\documentclass[journal=ACS Phot,layout=onecolumn, manuscript=article]{achemso}

\def\GG{{\overleftrightarrow{{\bf G}}}}

\usepackage{geometry}
\usepackage[font=small]{caption}

\usepackage[version=3]{mhchem} % Formula subscripts using \ce{}
\usepackage{lmodern} 
\usepackage{xcolor}
\usepackage{hyperref}
\usepackage{microtype}
\usepackage{lipsum}
\usepackage{amsmath,amssymb} 
\usepackage{dcolumn}
\usepackage{braket}

\SectionNumbersOn

\author{Álvaro Buendía}
\affiliation{Instituto de Estructura de la Materia, Consejo Superior de Investigaciones Científicas, Serrano 121, 28006 Madrid, Spain}
\email{a.buendia@csic.es}
\author{Jose A. Sánchez-Gil}
\affiliation{Instituto de Estructura de la Materia, Consejo Superior de Investigaciones Científicas, Serrano 121, 28006 Madrid, Spain}
\author{Vincenzo Giannini}
\affiliation{Instituto de Estructura de la Materia, Consejo Superior de Investigaciones Científicas, Serrano 121, 28006 Madrid, Spain}
\altaffiliation{Technology Innovation Institute, Masdar City 9639, Abu Dhabi, United Arab Emirates}
\alsoaffiliation{Centre of Excellence ENSEMBLE3 sp. z o.o., Wolczynska 133, Warsaw, 01-919, Poland}
\author{William L. Barnes}
\affiliation{Department of Physics and Astronomy, University of Exeter, Stocker Road, Devon, EX4 4QL, UK}
\author{Marie S. Rider}
\affiliation{Department of Physics and Astronomy, University of Exeter, Stocker Road, Devon, EX4 4QL, UK}

\email{m.s.rider@exeter.ac.uk}

\title{Long-range molecular energy transfer \\ mediated by strong coupling to \\ plasmonic topological edge states}

\begin{document}

% Abstract
\abstract{Strong coupling between light and molecular matter is currently attracting interest both in chemistry and physics, in the fast-growing field of molecular polaritonics. The large near-field enhancement of the electric field of plasmonic surfaces and their high tunability make arrays of metallic nanoparticles an interesting platform to achieve and control strong coupling. Two dimensional plasmonic arrays with several nanoparticles per unit cell and crystalline symmetries can host topological edge and corner states. Here we explore the coupling of molecular materials to these edge states using a coupled-dipole framework including long-range interactions. We study both the weak and strong coupling regimes and demonstrate that coupling to topological edge states can be employed to enhance highly-directional long-range energy transfer between molecules.
 }

\maketitle

% Section: Introduction 
\section{Introduction}
\label{sec:introduction}

In recent years, the strong interaction of light and molecular materials has attracted a growing wave of interest, due to possible applications in photonics, photo-chemistry and molecular quantum optics~\cite{Torma2015,ribeiro2018polariton,dovzhenko2018light,nikolis2019strong,herrera2020molecular,polak2020manipulating,hirai2023molecular,ebbesen2023introduction}. Molecules and photonic modes are in the strong coupling regime when their interaction strength exceeds their individual losses. This leads to a hybridization of molecule and photon, producing polaritons, which are part-matter, part-light. Early work focused on closed cavities as the source of confined photonic modes, while more recent work has highlighted the merits of open photonic systems such as planar surfaces, gratings and nanoparticle arrays. In systems such as these, the molecular material is directly accessible - crucial for direct measurement - whilst supporting photonic modes that are generally dispersive or highly delocalised, which can be exploited for enhancing energy transfer between molecules strongly coupled to the photonic modes~\cite{feist2015extraordinary,Du2018,georgiou2021ultralong,balasubrahmaniyam2022unveiling,jin2023enhanced,engelhardt2023polariton}. 

In this work we focus on topological arrays of plasmonic nanoparticles~\cite{Rider2019, BlancodePaz2019}; many studies have explored the strong coupling of molecules to plasmonic modes, for example in single nanoparticles \cite{Lee2023,Hamza2023}, cavities \cite{Bitton2022,SanchezBarquilla2022}, gratings \cite{Rider2022}, surface lattice modes \cite{Vakevainen2014}, chains of nanoparticles \cite{Allard2021,Allard2022, Allard2023}, and single topological plasmonic nanoparticles \cite{Thanopulos2023}.  However, the combination of topological states and collective strong coupling has yet to be explored to its full potential, although the coupling of a single quantum emitter to a topological photonic mode has been successfully demonstrated~\cite{barik2018topological}. Here we present a theoretical study of molecules coupled to the topological edge states of a metallic nanoparticle array, with the aim of enhancing and protecting directional molecular energy transport.  

Using a coupled-dipole framework~\cite{Rider2022, Kim2023}, we have developed the tools to study strong coupling between molecules and arrays of nanoparticles. This model can be applied to study any system which can be described as an heterogeneous array of coupled dipoles. 

In Section~\ref{sec:topo_edge_states_finite_array}, we study the dispersion bands and optical properties of the 2D Su-Schrieffer-Heeger (SSH) plasmonic array and its edge states. In Section~\ref{subsec:weak_ET}, we calculate the molecular energy transfer enhancement in the weak coupling regime. In Section~\ref{subsec:strong_ET} we study the strong coupling between molecules and edge states and strong-coupling enhanced molecular energy transfer. We demonstrate this by considering molecules as individual dipoles in the coupled-dipole model, and verify this result by also using a continuum model in which the molecular material is modelled as an effective medium. 

% Section: Topological edge states in a finite array of plasmonic nanoparticles
\section{Topological edge states in a finite array of plasmonic nanoparticles} 
\label{sec:topo_edge_states_finite_array}

 The simplest model showing topological edge states is the Su-Schrieffer-Heeger (SSH) model, initially introduced to explain the physics of polyacetalyene~\cite{su1979solitons}. The SSH model is a tight binding 1D-model with two staggered nearest-neighbours hoppings, $t_1$ and $t_2$, which represent the single and double bonds in the polymer. When terminated at the strong bond, the system hosts 0D edge states protected by the chiral symmetry of the lattice. The SSH model is applicable to a wide range of analogue systems, from radio-frequency metamaterials \cite{Baraclough_ACSPhot_2019_6_3003} to plasmonics. Furthermore, the plasmonic analogue of the SSH model~\cite{Downing2017} offers much complexity and richness, due to non-Hermiticity, all-to-all interactions~\cite{Pocock2018} and tunable polarization~\cite{Buendia2023}.

\begin{figure}[h!]
\begin{center}
\includegraphics[width=0.99\textwidth]{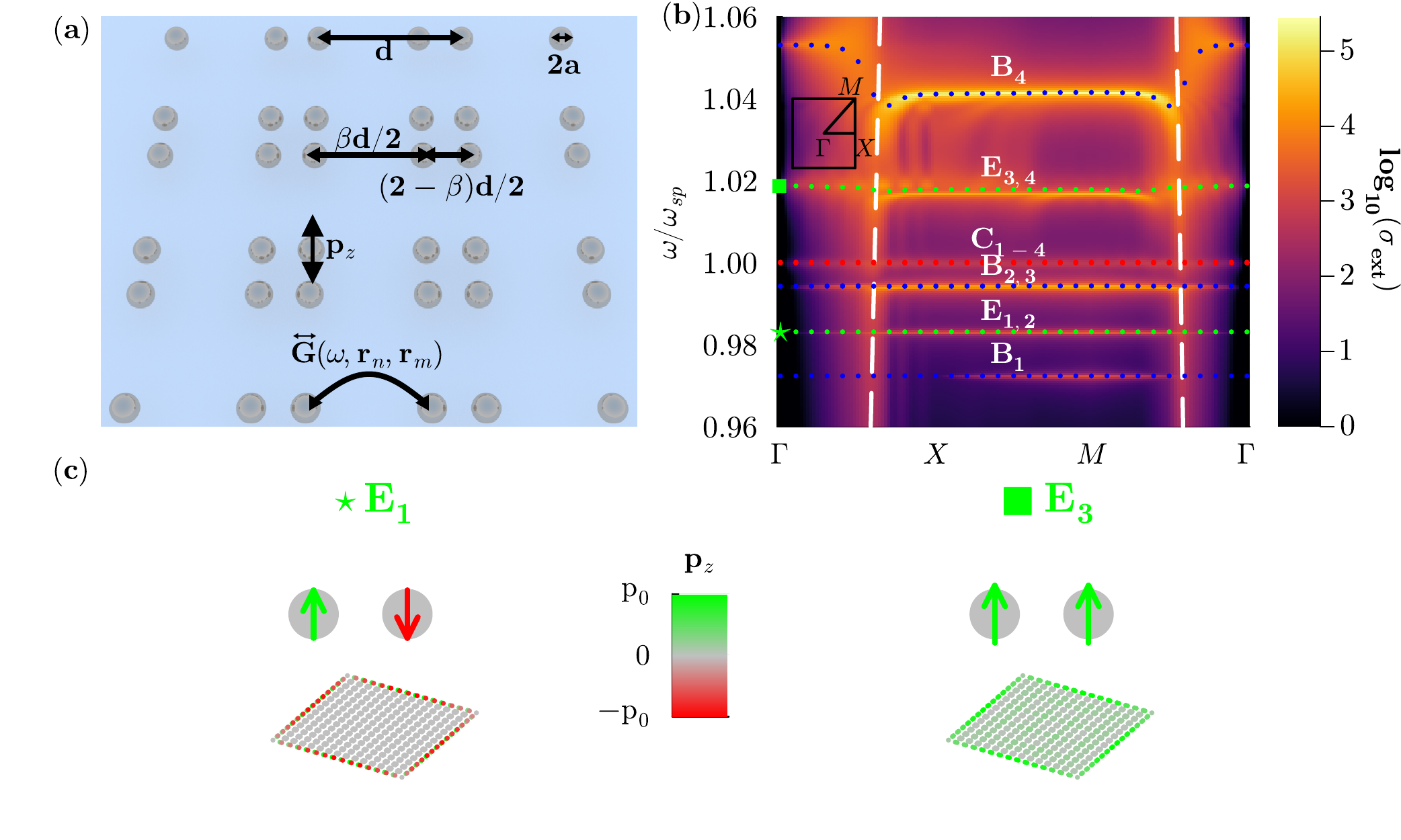}
\end{center}
\caption{\textbf{SSH2D array of metallic nanoparticles}: (a) Scheme for a finite $6\times 6$ SSH2D array. (b) Bulk ($B_1-B_4$, blue dotted lines), edge ($E_1-E_4$, green dotted lines)  and corner ($C_1-C_4$, red  dotted lines) dispersion bands of an array of $30\times 30$ metallic nanoparticles, with $a = 10$~nm, $d = 150~$nm, intra-cell distance $\beta d/2 = 0.8d$ and $\hbar\omega_{\mathrm{sp}} =2.50$~eV and losses $\hbar\gamma = 1$~meV, overlaid on top of optical extinction $\sigma_{\mathrm{ext}}$. (c) Main edge state ($E_1,E_3$) mode contributions at $\Gamma$,  where all the dimers over a single edge oscillate in-phase.}
\label{fig:SSH2D}
\end{figure}

The SSH2D model (see scheme in Fig.~\ref{fig:SSH2D}(a)) is the two-dimensional extension of the SSH model. The particles are ordered in a square lattice, with alternating distances $\beta d/2$ and $(2-\beta)d/2$ in the $x$ and $y$ directions, where $d$ is the lattice parameter and $\beta$ is a dimensionless parameter that ranges from 0 to 2. For $\beta > 1$ the array is in the topologically non-trivial phase and hosts both 1D (edge) and 0D (corner) states. 

2D plasmonic arrays of nanoparticles with several particles per unit cell and rotational symmetries have been shown to host topological edge and corner states~\cite{Honari2019, Proctor2019,Kim2020, Kim2020lasing, Rider2022}, which are robust to disorder and perturbations. Topological corner states have recently been proposed for lasing applications~\cite{Wu2023, Kim2020lasing}, while the quasi-1D character of the edge states makes them suitable for energy transfer applications. 

The edge states in the SSH2D model~\cite{Xie2018,Obana2019, Benalcazar2020, Kim2020, Schlomer2021, Luo2023, Liu2019, Hu2021, Heilmann2022} are protected by weak symmetries, namely the (generalized) sublattice symmetry and rotational ($C_4$) and mirror symmetries - these do not offer robustness against backscattering, however, the position in the unit cell and circular polarization of the emitter can impose handedness in the transport~\cite{Proctor2019}. Coupling achiral plasmonic modes to chiral molecules can also lead to chiral behaviour, as Ref.~\cite{Goerlitzer2023, Xiao2023} showed for surface lattice resonances. 

When metallic nanoparticles are small and roughly spherical they can be described as single electric dipoles, $\textbf{p} = \overleftrightarrow{\alpha}(\omega) \textbf{E}$, where $\overleftrightarrow{\alpha}(\omega)$ is the electric polarizability of the nanoparticle and $\textbf{E}$ is the incident electric field. For a single nanosphere, the polarizability tensor acts like a scalar. In the quasi-static limit, i.e. when the particle is small compared to the wavelength of the light ($a\ll \lambda$), the polarizability $\alpha_{\mathrm{QS}}(\omega)$ is: 
\begin{equation}
\alpha_{\mathrm{QS}}(\omega) = 4\pi a^3 \epsilon_0 \frac{\epsilon(\omega) - \epsilon_\mathrm{B}}{\epsilon(\omega) + 2\epsilon_\mathrm{B}},
\label{eq:alphaqs}
\end{equation}
where $\epsilon_B$ is the background permittivity and $\epsilon(\omega)$ the electric permittivity of the metal, which in the Drude model is:
\begin{equation}
\epsilon(\omega) = \epsilon_{\infty} - \frac{\omega_p^2}{\omega(\omega+i\gamma)}.
\end{equation}
\label{eq:permittivity}
where $\epsilon_\infty$ is the static dielectric constant, $\omega_p$ is the plasma resonance and $\gamma$ is the optical loss of the metal. 

The quasi-static polarizability neglects damping, so we make the radiative correction: 
\begin{equation}
\alpha(\omega) = \frac{\alpha_{\mathrm{QS}}(\omega)}{1-i\frac{k^3}{6\pi}\alpha_{\mathrm{QS}}(\omega)}. 
\end{equation}
\label{eq:alpha}

 Metallic nanospheres show an optical resonance at the frequency $\omega_{\mathrm{sp}} = \omega_{p}/\sqrt{2\epsilon_B+\epsilon_\infty}$, known as localized surface plasmon resonance (LSPR).

When instead of a single metallic nanoparticle we employ an ensemble of them, the plasmonic modes of the particles hybridize, leading to collective plasmonic modes. When the distances between the particles are at least $3a$, the dipolar approximation holds and the array can be described as a system of coupled dipoles. 

Coupled electric dipole models are routinely employed to study periodic arrays of plasmonic nanoparticles \cite{Cherqui2019, Zundel2022}. However, edge states are not present in the infinite periodic array as they arise from the breaking of translational symmetry, and so it is necessary here to work with finite arrays. For an array of $N$ nanoparticles, we have a system of $N$ coupled-dipole equations:
\begin{equation}
\overleftrightarrow{\alpha}^{-1}(\omega)\textbf{p}_n =  \frac{k^2}{\epsilon_0} \sum_{m\neq n} \overleftrightarrow{\textbf{G}}(\omega, \textbf{r}_n, \textbf{r}_m) \textbf{p}_m,
\label{eq:coupleddip}
\end{equation}
where $\overleftrightarrow{\textbf{G}}(\omega, \textbf{r}_n, \textbf{r}_m)$ is the Green dyadic's function, which mediates the interaction between two dipoles. The Green dyadic function is given by:
\begin{equation}
\begin{split}
\GG(\omega,\textbf{r}_n, \textbf{r}_m) &= \frac{e^{ikR}}{4\pi R}\bigg[\left(1 + \frac{i}{kR} - \frac{1}{k^2 R^2}\right)\mathcal{I}  - \left(1 + \frac{3i}{kR}-\frac{3}{k^2 R^2}\right)\frac{\textbf{R}\otimes \textbf{R}}{R^2} \bigg] ,
\end{split}
\label{eq:Greensdyad}
\end{equation}
where $\mathcal{I}$ is the identity matrix, $\textbf{R} = \textbf{r}_m - \textbf{r}_n$ and $R=|\textbf{R}|$. 

In the near-field ($kR \ll 1$), $\GG(\omega, \textbf{r},\textbf{r}_0)$ can be approximated by the quasi-static Green dyadic:

\begin{equation} 
\begin{split}
\GG_{\mathrm{QS}}(\omega,\textbf{r}_n, \textbf{r}_m) &= \frac{1}{4\pi k^2 R^3}\bigg[-\mathcal{I}
+ 3 \frac{\textbf{R}\otimes \textbf{R}}{R^2} \bigg].
\end{split}
\label{eq:QSGreensdyad}
\end{equation}

However, we will consider the full Green dyadic and long-range interactions in the array and will simply use this approximation to understand near-field coupling. 

The coupled-dipole system of equations (Eq. \eqref{eq:coupleddip})  can be simply rewritten in a matricial way as:
\begin{equation}
\lambda_i(\omega_i) = 0,
\label{eq:eigvals}
\end{equation}
where $\lambda_i$ is the $i$-th eigenvalue of $\overleftrightarrow{\alpha}_{\mathrm{eff}}^{-1}(\omega)$, which is the inverse of the effective polarizability of the array, whose $3\times3$ blocks are given by:
\begin{eqnarray}
[\overleftrightarrow{\alpha}_{\mathrm{eff}}(\omega)]^{-1}_{3n-2;3n;3m-2;3m} =  \begin{cases}\overleftrightarrow{\alpha}^{-1}(\omega)  & \mbox{if } n = m,
 \\
\frac{k^2}{\epsilon_0}\overleftrightarrow{ \textbf{G}}(\omega, \textbf{r}_n, \textbf{r}_m) & \mbox{if } n \neq m.
\end{cases}
\label{eq:alpha_eff}
\end{eqnarray}
The dipolar moments of the normal modes are given by $\textbf{P}_i = \begin{pmatrix}
\textbf{p}_1 \dots  ,\textbf{p}_N\end{pmatrix}^T$, which is the $i$-th eigenvector of $\overleftrightarrow{\alpha}^{-1}_{\mathrm{eff}}(\omega_i)$.  

When periodicity is broken, the normal modes no longer have a defined momentum $\textbf{k}$, so the dispersion bands $\omega(\textbf{k})$ are not defined. The usual way to obtain the dispersion bands in systems with broken periodicity is the supercell method \cite{Kim2023}, where a larger ancillary unit cell is used to recover periodicity. However, long-range interactions in periodic arrays imply infinite lattice sums, and for large unit cells this approach would be computationally very consuming and so we take a different route. Instead, we use the spatial dependence of the eigenmodes to infer the dependence on $\textbf{k}$. By classifying the eigenmodes $\textbf{P}_i$ by symmetries (see Appendix \ref{sec:NormalModeSymmetries} for more information) and localization (bulk, edge and corner states), we obtain effective dispersion bands for a finite array: 
\begin{equation}
\omega_{\mathrm{sym}}(\textbf{k}) = \frac{\sum_{i = \{i_{\mathrm{sym}}\}} \omega_i |\textbf{P}_{\textbf{k},i}|^2}{\sum_{i = \{i_{\mathrm{sym}}\}} |\textbf{P}_{\textbf{k},i}|^2},
\label{eq:Dispersion}
\end{equation}
where $\textbf{P}_{\textbf{k},i}= \textbf{P}_i \cdot \begin{pmatrix}
e^{i\textbf{k}\textbf{r}_1}  \cdots  e^{i\textbf{k}\textbf{r}_N} \end{pmatrix}^T$.

In Fig.~\ref{fig:SSH2D}(b) we plot the dispersion bands for the out-of-plane ($\textbf{p}_z$) polarization. This corresponds to considering only $zz$ terms in previous tensorial equations. We choose out-of-plane polarizations for the sake of simplicity, as $xx$ and $yy$ modes are hybridized, which duplicates the number of bands to study. Additionally, the out-of-plane polarization respects all the spatial symmetries of the SSH2D lattice, which facilitates the existence of corner and edge states. 

Out-of-plane bands can be spectrally isolated from in-plane bands using elongated particles (e.g. spheroids) with the long axis oriented parallel to $\textbf{u}_z$ \cite{Proctor2020}. Here, for simplicity, we ignore the in-plane contribution. We choose a SSH2D array of $15\times 15$ unit cells (i.e. $30 \times 30$ nanoparticles), with parameters $a = 10$~nm, $d = 150$~nm, $\beta=1.6$ and $\hbar\omega_{\mathrm{sp}} =2.50 eV$ with losses $\hbar\gamma = 1$~meV. We consider reduced losses compared to silver or gold, as large losses can make edge, bulk and corner states overlap spectrally. This approximation allows for a clear analysis of the phenomena described in this work, but should be noted when comparing to an experimental system using common metals. We choose the largest $\beta$ within the dipolar approximation limits $(2-\beta)d/2 \geq 3a $ to maximize gaps and avoid this overlapping.
Red, green and blue lines represent, respectively, corner, edge and bulk bands. Light lines are shown as white dashed lines. 

In Fig.~$\ref{fig:SSH2D}$(b) the dispersion bands are overlaid on top of the optical extinction, which is given by~\cite{Proctor2019}: 
\begin{eqnarray}
\sigma_{\mathrm{ext}}(\omega,\textbf{k}) =  4\pi k \Im\left(\frac{\hat{\textbf{E}}_{\mathrm{inc}}^*\overleftrightarrow{\alpha}_{\mathrm{eff}}(\omega)\hat{\textbf{E}}_{\mathrm{inc}}}{N |\textbf{E}_0|^2}\right),
\label{eq:SigmaExt}
\end{eqnarray}
where $\hat{\textbf{E}}_{\mathrm{inc}}$
is the incoming electric field (a plane wave) at the positions of the nanoparticles, i.e. $\left[\hat{\textbf{E}}_{\mathrm{inc}}\right]_{3n-2, 3n}= \textbf{E}_0e^{i(\textbf{k}\textbf{r}_n-\omega t)}$. We consider both propagating and evanescent solutions with real or imaginary $k_z = \sqrt{k^2-k_x^2-k_y^2}$ to probe the whole reciprocal space. We see the upper bands are brighter and spectrally broader than the lower frequency bands. This is due to the symmetries of the bands (see Appendix \ref{sec:NormalModeSymmetries} for more information), as dipoles within a single unit cell in $B_4$ and $E_{3,4}$ states are in-phase, their emitted fields interfere constructively in the far-field. 

The plasmonic bands, specially the lower ones, are nearly flat. This is due to long-range interactions and to the clustering of nanoparticles as $\beta \rightarrow 2$, which become quadrumers in the bulk, dimers along the edge, and monomers in the corners.

The edge states are doubly degenerate, and disperse similarly to the bulk states of 1D SSH chains~\cite{Schlomer2021}. For the upper gap edge states, the particles in dimers oscillate in phase. These are symmetric or bright edge states. However for the lower gap edge states, the particles in each dimer oscillate out of phase. These correspond to antisymmetric or dark edge states. In Fig.~\ref{fig:SSH2D}(c) we represent the main contribution at $\Gamma$ for a dark ($E_1$) and a bright ($E_3$) edge  state, i.e. the modes where the dimers along a single edge oscillate in phase.

The corner state bands are flat and quadruply-degenerate, and are fixed at  $\omega_{\mathrm{sp}}$ due to the generalized sublattice symmetry~\cite{Luo2023}.

The notation of dark and bright bands refers to the extent to which a particular mode will couple to a plane wave; dark modes can in fact be excited and probed by near-field local coupling~\cite{Abujetas2021}, despite their name. In our system the dark bands are particularly of interest due to their flatness, as flat bands can pose a large local density of states (LDOS). To derive the LDOS of the metasurface~\cite{Lunnemann2016, Abujetas2021}, first we introduce an effective Green function  $\overleftrightarrow{\textbf{G}}_{\mathrm{eff}}(\omega,\textbf{r}, \textbf{r}_{0})$ that represents the interaction between two dipoles at position $\textbf{r}$ and $\textbf{r}_{0}$ via excitation of the array of plasmonic nanoparticles:
\begin{eqnarray} \overleftrightarrow{\textbf{G}}_{\mathrm{eff}} (\omega, \textbf{r}, \textbf{r}_{0})   = \frac{k^2}{\epsilon_0} \sum_{i,j=1}^{N} 
\overleftrightarrow{\textbf{G}}(\omega, \textbf{r}, \textbf{r}_m) 
\left[\overleftrightarrow{\alpha}_{\mathrm{eff}}(\omega)\right]_{nm}
\overleftrightarrow{\textbf{G}}(\omega, \textbf{r}_n, \textbf{r}_{0}).
\label{eq:EffectiveGreensDyad}
\end{eqnarray}
The local density of states at position $\textbf{r}_0$, in terms of the effective Green function, is then \cite{Lunnemann2016}:
\begin{equation}
\frac{\rho(\textbf{r}_0, \omega)}{\rho_0} = 1 +  \frac{2\pi}{k} \Im \bigl\{ \operatorname{Tr}\left[\overleftrightarrow{\textbf{G}}_{\mathrm{eff}}(\omega, \textbf{r}_{0},\textbf{r}_{0})\right]\bigr\} ,
\label{eq:LDOS}
\end{equation}

where $\rho_0 = \omega^2/\pi c^3$ is the density of electromagnetic modes in blackbody radiation. In order to calculate the LDOS, we choose an out-of-plane polarization for the probe too, as the near-field dipole-dipole interaction (EQ.~\ref{eq:QSGreensdyad}) between the probe and the $n$th nanoparticle will be maximum when their dipoles are parallel and oriented in the direction of the vector between their positions $\textbf{R} = \textbf{r}_n - \textbf{r}_0$. The electric field hotspots, and therefore the maxima of LDOS, are expected to be over nanoparticles, where $\textbf{R} = R \textbf{u}_z$. 

This is another justification to focus on the out-of-plane modes of the plasmonic array, as the hotspots of in-plane modes will be in-plane too ($z=0$), which make them more challenging to probe and couple to.

While the frequency of the probe will generally be fixed, the plasmonic array resonance frequencies can be fine-tuned by changing the size, shape and material of the nanoparticles and the lattice parameters. However, since we are not trying to simulate the response of any specific real molecule, we fix the plasmonic array parameters and vary the frequency of the probe $\omega_{mol}$. 
\begin{figure}[H]
\begin{center}
\includegraphics[width=0.935\textwidth]{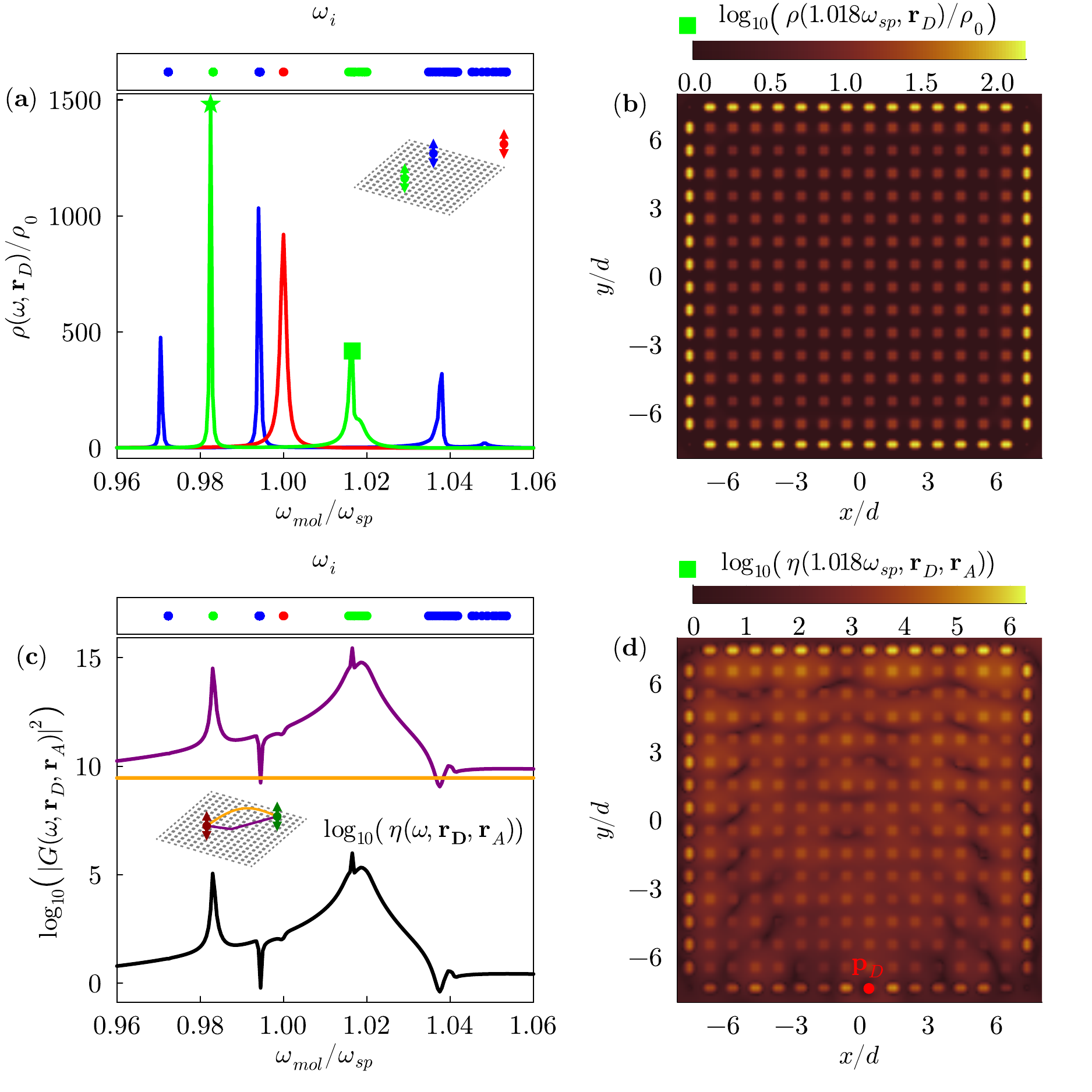}
\end{center}
\caption{\textbf{Local density of states and energy transfer enhancement via weak coupling to dark and bright edge states}: (a) Local density of states for a $30\times 30$ array of metallic nanoparticles, with $a = 10$~nm, $d = 150$~nm, intra-cell distance $\beta d/2 = 0.8d$, $\hbar\omega_{\mathrm{sp}} =2.50$~eV and losses $\hbar\gamma = 1$~meV. LDOS varies with $\omega$ for out-of-plane polarization, and is given for three instances of $\mathbf{r}_\mathrm{D}$: over the edge (green line), bulk (blue line), and corner (red line) at $z=0.2d$.  (b) Local density of states of the same array at the frequency of the bright edge state $\omega = 1.018\omega_{\mathrm{sp}}$. The local maxima of the LDOS match with the spatial positions of the nanoparticles. The absolute maxima are at the edge, as expected. (c) Square of the free-space Green function of two molecules over adjacent edges (orange line) versus square of the effective Green function (purple line), mediated by coupling to the plasmonic array. Black line represents the enhancement of the energy transfer. Energy transfer is enhanced for almost all frequencies, but especially near the frequency of the bright edge state. (d) Energy transfer enhancement at the same frequency for donor fixed at $\mathbf{r}_\mathrm{D}$ and mapping over the position of acceptor $\mathbf{r}_\mathrm{A}$. The energy transfer between donor and acceptor is maximally enhanced along the edge due to coupling with the edge state.}
\label{fig:weakcouplingdark}
\end{figure}

In Fig.~\ref{fig:weakcouplingdark}(a), red, green and blue lines represent the LDOS over a nanoparticle at the corner, edge and bulk, at a height $z=0.2d$. We can see the peaks match with the frequencies of the normal modes, plotted under the LDOS. We see the maximum LDOS is at the frequency of the dark edge states.
In panel (b) we plot the LDOS at the frequency of the bright edge state $(\omega_{mol}=1.018\omega_{sp})$, mapping the position of the probe $\textbf{r}_0$ at constant height $z=0.2d$. We see the maxima of the LDOS are over the positions of the nanoparticles on the edge of the array, as expected.

\section{Enhanced molecular energy transfer in the presence of topological edge states}
\label{sec:ET}

In this section, we discuss the effect of the topological plasmonic array on the energy transfer between two molecules located near the array.  Molecular energy transfer mediated by the dipole-dipole interaction, known as Förster resonance energy transfer (FRET), is proportional to the square of the Green's dyadic ~\cite{Novotny2012}, i.e., at short distances, it decays like $R^{-6}$. In the medium and long-range energy transfer decays like $R^{-4}$ and $R^{-2}$, respectively, but in these medium and long-range regimes energy transfer is not experimentally significant. To extend the range of the energy transfer for molecular spectroscopy or the catalysis of chemical reactions, the molecules can be coupled to plasmonic modes, due to their long-range nature ~\cite{Hamza2023}. 

In Section~\ref{subsec:weak_ET} we study the system of a single donor molecule and a single acceptor molecule, spatially separated and located above the array. We demonstrate that by choosing array parameters such that the donor and acceptor molecules have resonant frequency commensurate with the topological edge state, molecular energy transfer is both greatly enhanced with respect to free-space energy transfer, and also highly directional along the edge of the system. In Section~\ref{subsec:strong_ET} we move on to study the more realistic system of a molecular ensemble on top of the nanoparticle array. We demonstrate that this system may enter the strong coupling regime, in which the enhancement of molecular energy transfer via coupling to a topological edge state may be increased even further via the process known as polariton assisted resonance energy transfer (PARET)~\cite{Du2018}.

\subsection{Energy transfer enhancement via weak coupling between molecules and topological edge states}
\label{subsec:weak_ET}

From now on, we consider the molecules as electric dipoles. Interaction involving higher multipoles are very short-ranged so the main contribution to energy transfer will be the dipole-dipole contribution. The power transmitted from the donor molecule dipole at position $\mathbf{r}_\mathrm{D}$ to an acceptor molecule dipole at position $\mathbf{r}_\mathrm{A}$ is given by \cite{Novotny2012}: 
\begin{equation}
P_{\mathrm{D}\rightarrow \mathrm{A}} = \frac{\omega}{2} \Im(\alpha_{\mathrm{mol}}(\omega)) \Big|\textbf{u}_\mathrm{A} \cdot \textbf{E}_\mathrm{D}(\textbf{r}_\mathrm{A})\Big|^2 ,
\label{eq:EnergyTransfer}
\end{equation}
where  $\textbf{u}_\mathrm{A}$ is a unitary vector in the direction of the acceptor dipole, $\alpha_{\mathrm{mol}}(\omega)$ is the polarizability of the molecules, which we model as a Lorentzian centered at frequency $\omega_{\mathrm{mol}}$ and width $\Gamma$:
\begin{equation}
\alpha_{\mathrm{mol}}(\omega) = - \frac{6\pi\epsilon_0}{k^3} \frac{\Gamma/2}{(\omega-\omega_{\mathrm{mol}})-i\Gamma/2},
\label{eq:alpha_mol}
\end{equation}
and $\textbf{E}_\mathrm{D}(\mathbf{r}_\mathrm{A})$ is the electric field created by the donor dipole at position $\textbf{r}_\mathrm{A}$:
\begin{align}
    \mathbf{E}_\mathrm{D}(\mathbf{r}_\mathrm{A}) &= \frac{k^2}{\epsilon_0 \epsilon_\mathrm{B}}\overleftrightarrow{\mathbf{G}}(\omega, \mathbf{r}_\mathrm{A},\mathbf{r}_\mathrm{D})\mathbf{p}_\mathrm{D}.
    \label{eq:Efield_GreensFunction}
\end{align}

In the presence of the plasmonic array, the power transferred via excitation of the array is given by:
\begin{equation}
P_{\mathrm{D}\rightarrow \mathrm{A}} = \frac{\omega_{mol}}{2} \Im(\alpha_{\mathrm{mol}}(\omega_{mol})) \Big|\textbf{u}_\mathrm{A} \cdot \overleftrightarrow{\textbf{G}}_{\mathrm{eff}}(\omega_{mol}, \textbf{r}_\mathrm{A}, \textbf{r}_\mathrm{D})\textbf{p}_\mathrm{D}\Big|^2 
\label{eq:EnhancementEnergyTransfer}
\end{equation}

We define an enhancement factor of the power transmitted from donor to acceptor by the presence of the array:
\begin{equation}
\eta(\omega_{mol}, \textbf{r}_\mathrm{D}, \textbf{r}_\mathrm{A})= \frac{\Big|\mathbf{u}_\mathrm{A} \cdot\overleftrightarrow{\mathbf{G}}_{\mathrm{eff}}(\omega_{\mathrm{mol}}, \mathbf{r_\mathrm{A}},\mathbf{r}_\mathrm{D})\mathbf{u}_\mathrm{D}  \Big|^2}{\Big| \mathbf{u}_\mathrm{A} \cdot \overleftrightarrow{\mathbf{G}}(\omega_{\mathrm{mol}}, \mathbf{r}_\mathrm{A},\mathbf{r}_\mathrm{D})\mathbf{u}_\mathrm{D}\Big |^2},
\label{eq:enhancementfactorweak}
\end{equation}

where $\textbf{u}_D$ is a unitary vector in the direction of the donor dipole. In Fig.~\ref{fig:weakcouplingdark}(c) we plot the square of the effective Green function via the array (purple line) versus the square of the free-space dipole-dipole Green function (orange line) for the donor and acceptor molecules positioned over adjacent edges. We also plot the enhancement factor, which is the quotient of the two.  We see that energy transfer is enhanced for the entirety of the interval $\omega_{mol} \in (0.96\omega_{sp}, 1.06\omega_{sp})$, but that enhancement is maximum when the molecular resonance is close to the frequencies of the edge states. There are two dips in the energy transfer enhancement around $0.995\omega_{sp}$ and $1.035\omega_{sp}$, matching with resonance frequencies of bulk states, which might prevent coupling to edge states.

Even though the dark edge states exhibit a larger LDOS, the energy transfer enhancement via the bright edge states is larger by almost a factor $10^2$. This may be due to destructive interference in the electric field radiated by the dark state. However, the peak in the energy transfer for the dark edge state is narrower, as dark states are lower in loss and less screened by nearby bulk and corner states. 

In Fig.~\ref{fig:weakcouplingdark}(d) we map the energy transfer enhancement at the frequency of the bright edge state $(\omega = 1.018\omega_{\mathrm{sp}})$. We fix the position of the donor over a hotspot, or a maximum of the LDOS,
and vary the position of the acceptor. We see that the energy transfer is strongly enhanced along the edge. 

\subsection{Energy transfer enhancement via strong coupling between molecules and topological edge states}
\label{subsec:strong_ET}

In a realistic molecule-nanophotonic system, the molecular material will generally contain a great number of molecules, $N_\mathrm{mol}$, instead of the simple picture of two isolated donor and acceptor molecules. In fact, it is the behaviour of molecular ensembles with $N_\mathrm{mol}$ large enough to enter the strong coupling regime that are at the centre of the burgeoning fields of molecular polaritonics and polaritonic chemistry~\cite{Ebbesen_ACS_Accounts_2016_49_2403,herrera2020molecular}. While in Section~\ref{subsec:weak_ET} we demonstrated the ability of topological edge states to enhance molecular energy transfer via weak coupling, we now describe two complementary methods to study a system of molecules strongly coupled to modes of the topological array. 

When a plasmonic mode couples to  an emitter, two hybridized bands form, lower ($\omega_\mathrm{{LP}}<\omega_{\mathrm{mol}}$) and upper ($\omega_{\mathrm{UP}} > \omega_{\mathrm{mol}}$) polaritons, separated by a spectral gap ($\Omega = \omega_{\mathrm{UP}} - \omega_{\mathrm{LP}}$) 
which is known as Rabi splitting. When this gap exceeds the plasmonic and molecular losses of the original uncoupled modes, the system is strongly coupled. 

We first demonstrate that strong coupling to a topological edge state can be achieved by studying a moderate number of molecules placed in the field maxima above the nanoparticles at the system edge. Using this model, we demonstrate strong coupling to a topological edge state and subsequent enhancement of molecular energy transfer. Then, in order to consider a system of realistic molecular density and spatial positioning, we treat the molecular system as an effective continuous medium in which the nanoparticle array is embedded. Using this effective medium model we then study the effect of strong coupling on molecular energy transfer. We show the results of the two method give good qualitative agreement.

\subsubsection{Demonstration of strong coupling to a topological edge state with a finite array of molecules}
\label{subsubsec:demo_ST}

The effective polarizability of the molecular array (analogous to Equation~\eqref{eq:alpha_eff} for the nanoparticle array) is:
\begin{eqnarray}
[\overleftrightarrow{\alpha}_{\mathrm{eff}}^{\mathrm{mol}}(\omega)]^{-1}_{3n-2;3n;3m-2;3m} =   \begin{cases} \overleftrightarrow{\alpha}_{\mathrm{mol}}^{-1}(\omega)  & \mbox{if } n = m, \\
-\frac{k^2}{\epsilon_0}\overleftrightarrow{\textbf{G}}(\omega, \textbf{r}_n, \textbf{r}_m) & \mbox{if } n \neq m .
\end{cases}
\label{eq:alpha_eff_mol}
\end{eqnarray}
From Equation~\ref{eq:alpha_eff} and \ref{eq:alpha_eff_mol}, we obtain the effective polarizability of the whole system as:  
\begin{eqnarray}
[\overleftrightarrow{\alpha}_{\mathrm{eff}}^{\mathrm{NP}\mbox{-}\mathrm{mol}}(\omega)]^{-1}  =   \begin{pmatrix} \overleftrightarrow{\alpha}_{\mathrm{eff}}^{-1}(\omega) && -\frac{k^2}{\epsilon_0} \mathcal{G}^{\mathrm{NP}\mbox{-}\mathrm{mol}}(\omega)  \\ -\frac{k^2}{\epsilon_0}\left(\mathcal{G}^{\mathrm{NP}\mbox{-}\mathrm{mol}}(\omega)\right)^T && \overleftrightarrow{\alpha}_{\mathrm{eff}}^{\mathrm{mol}-1}(\omega) \end{pmatrix}
\label{eq:alpha_eff_NP-mol}
\end{eqnarray}

where $\mathcal{G}^{\mathrm{NP}\mbox{-}\mathrm{mol}}(\omega)$ is a $N\times N_{\mathrm{mol}}$ matrix whose blocks are given by the Green dyadic between nanoparticles and molecules.

From Equation~\ref{eq:alpha_eff_NP-mol} we can understand the coupling between molecular and plasmonic arrays. In the limit where the nanoparticle and molecular array polarizabilities do not spectrally overlap, both arrays are fully uncoupled. When the polarizabilities have spectral overlap and the Green function between molecules and nanoparticles is large enough, the system will be strongly coupled.

We expect the coupling to be optimal when the spectral and spatial overlap of the molecules with the edge states is maximal. In order to demonstrate strong coupling to edge states, we consider an array of molecules placed at a height $z$ above the nanoparticles at the edge of the system, coinciding with the spatial maxima of the LDOS as shown in Fig.~\ref{fig:weakcouplingdark}(b). 

\begin{figure}[H]
\begin{center}
\includegraphics[width=0.99\textwidth]{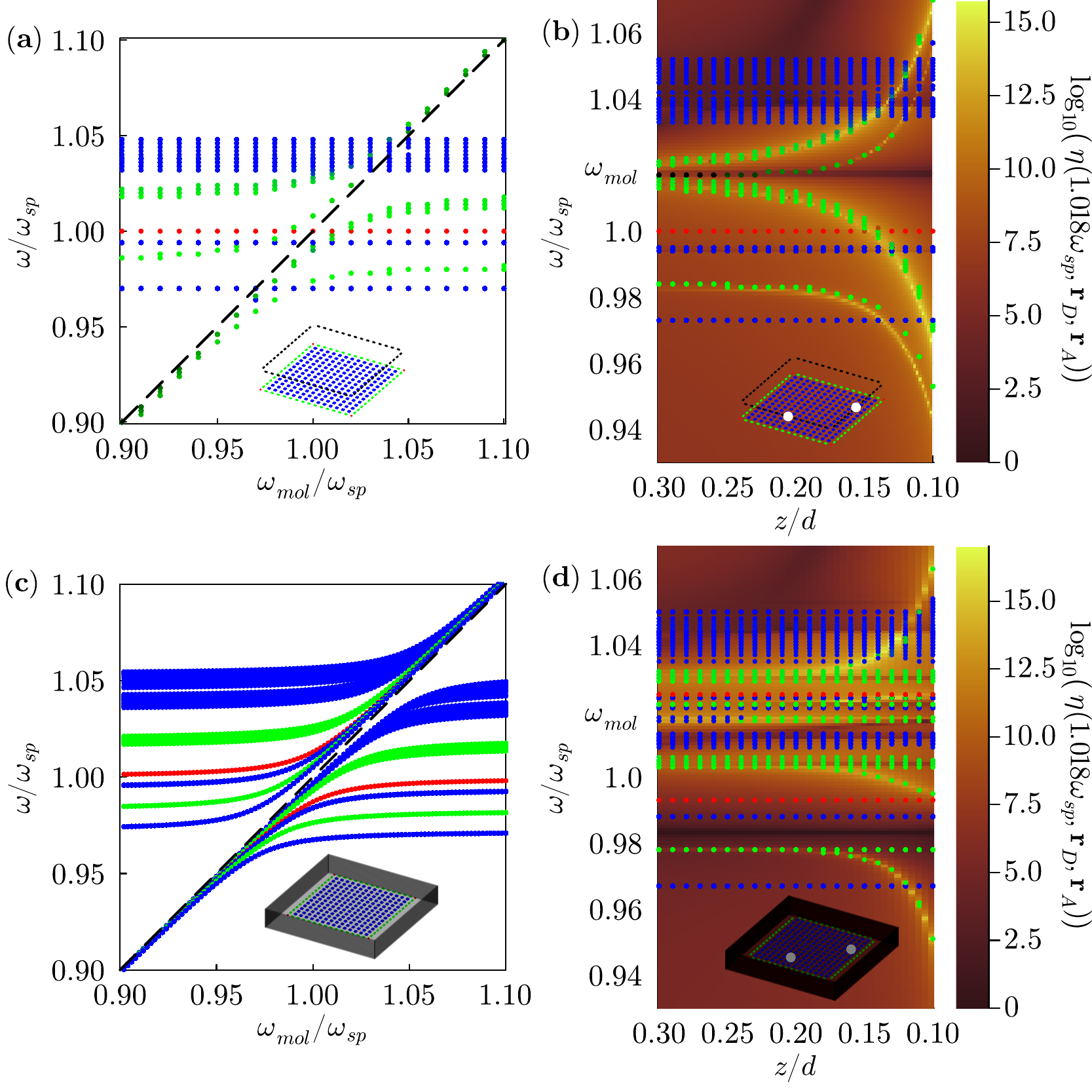}
\end{center}
\caption{\textbf{Energy transfer mediated by strong coupling between molecules and topological bright edge states}: (a) Strong coupling between array of molecules and plasmonic array of $30\times 30$ silver nanoparticles, with $a = 10nm$, $d = 150nm$, intra-cell distance $\beta d/2 = 0.8a$ and $\hbar\omega_{sp} = 2.50 eV$ and losses $\hbar\gamma = 1 meV$, depending on the molecular frequency $\omega_{mol}$. The molecules (black dots) are at the LDOS hotspots, over the nanoparticles on the edge. Red, green and blue dotted lines are corner, edge and bulk states. Black dashed line is $\omega = \omega_{mol}$. White dots are the donor/acceptor molecules (b) Spectrum of the strongly coupled system and energy transfer between molecules resonant with bright edge state ($\omega_{mol} = 1.018\omega_{sp}, \hbar\Gamma= 0.1meV$) on adjacent edges depending on the distance between the molecules and the plasmonic array $z/d$ overlaid with the energy transfer enhancement between molecules on adjacent edges.  (c) Strong coupling spectrum of the same plasmonic array embedded in an effective continuous molecular medium depending on its resonance frequency $\omega_{mol}$. (d) Spectrum of the strongly coupled system plasmonic array-effective medium and energy transfer between molecules resonant with bright edge state ($\omega_{mol} = 1.018\omega_{sp}, \hbar\Gamma= 0.1meV$) on adjacent edges depending on the distance between the molecules and the plasmonic array $z/d$.}
\label{fig:StrongCoup}
\end{figure}

In Fig.~\ref{fig:StrongCoup}(a) we plot the spectrum of the coupled system for $z=0.15d$ for varying $\omega_{\mathrm{mol}}$. We see each plasmonic edge band hybridize with the molecules and split into two polaritonic bands ($\omega_{UP}>\omega_{\mathrm{mol}}$ and $\omega_{LP}<\omega_{\mathrm{mol}}$). The corner and bulk modes however, only couple weakly to the molecules, due to their spatial distribution.

To compute the energy transfer enhancement, we define an effective Green function between any two molecules $n_{D},n_{A} \in [N+1, N+N_{mol}]$ in the molecular array via strong coupling to the plasmonic array. From  $\textbf{p} = \overleftrightarrow{\alpha}(\omega) \textbf{E}$ and Equation~\eqref{eq:Efield_GreensFunction}, we derive (see Appendix \ref{sec: Strongly_Coupled_Green_Dyadic}): 

\begin{eqnarray}
\overleftrightarrow{\textbf{G}}_{\mathrm{eff}}^{\mathrm{SC}}(\omega,\textbf{r}_\mathrm{A}, \textbf{r}_\mathrm{D})   = \frac{\epsilon_0}{k^2} \overleftrightarrow{\alpha}_{\mathrm{mol}}^{-1}(\omega)
 \cdot \left[\overleftrightarrow{\alpha}_{\mathrm{eff}}^{\mathrm{NP}\mbox{-}\mathrm{mol}}(\omega)\right]_{3n_A-2;3n_A,3n_D-2;3n_D}  \overleftrightarrow{\alpha}_{\mathrm{mol}}^{-1}(\omega)
\label{eq:SCEffGreen_1}
\end{eqnarray}

When $z\rightarrow\infty$
this effective Green function simply reduces to the free-space dipole-dipole Green function (see Appendix \ref{sec: Strongly_Coupled_Green_Dyadic}).

When the molecules are strongly coupled to the system, molecular energy transfer occurs via polariton modes - the molecular component of polaritons varies with frequency, resulting in a frequency-dependent energy transfer enhancement factor:
\begin{equation}
\eta(\omega, \omega_{\mathrm{mol}}, \textbf{r}_\mathrm{D}, \textbf{r}_\mathrm{A})= \frac{\Big|\mathbf{u}_\mathrm{A} \cdot\overleftrightarrow{\mathbf{G}}_{\mathrm{eff}}^{\mathrm{SC}}(\omega, \mathbf{r}_\mathrm{A},\mathbf{r}_\mathrm{D})\mathbf{u}_\mathrm{D}  \Big|^2}{\Big| \mathbf{u}_\mathrm{A} \cdot \overleftrightarrow{\mathbf{G}}(\omega_{\mathrm{mol}}, \mathbf{r}_\mathrm{A},\mathbf{r}_\mathrm{D})\mathbf{u}_\mathrm{D}\Big |^2}.
\label{eq:EnhancementEnergyTransferSC}
\end{equation}
In Fig.~\ref{fig:StrongCoup}(b) we plot the spectrum and energy transfer enhancement factor, varying the distance between nanoparticles and molecules from $z=0.3d$ to $z=0.1d$. Red, green and blue dots represent the frequencies of the corner, edge and bulk modes.

We choose the molecules to be resonant with the bright edge states, such that $\omega_{\mathrm{mol}} = 1.018\omega_{\mathrm{sp}}$, as these showed larger enhancement factors than the dark states in the weak coupling regime (as demonstrated in Section~\ref{subsec:weak_ET}). In Appendix C we study coupling to dark edge states. When the molecules are very close to the nanoparticle array ($z/d \lesssim  0.2$), they not only strongly couple with the edge state on resonance with the molecules, but also with the off-resonance edge states. This is due to plasmonic losses, meaning that the spectral overlap between edge states is non-zero.   

The colormap in Fig.~\ref{fig:StrongCoup}(b) represents the enhancement factor between two molecules over two particles in adjacent edges of the array.  As we see, the maxima of the enhancement matches with the frequencies of the strongly coupled edge states. We see the enhancement factor grows when $z$ decreases, i.e. it is proportional to the coupling strength. We see that for strongly coupled molecules we get factors up to $10^{15}$, contrasting with the maximum factors around $10^6$ in the weak coupling regime. 

\subsubsection{Strong coupling using a continuous effective medium model}
\label{subsubsec:continuous_effective_medium}

In a molecule-nanophotonic system made with straightforward fabrication techniques, molecules will not only be located at the hotspots above nanoparticles, nor only present over nanoparticles at the edge of the system. A more realistic system would entail a randomly distributed layer of molecules over the nanoparticle array, with a large number of molecules per unit cell. Modelling this molecular system via the same coupled-dipole approach as the previous section would be computationally exhaustive. In the limit that the average distance between molecules is much smaller than the distance between nanoparticles, spatial disorder will be negligible and we can treat the molecular system as an effective continuous medium in which the nanoparticles are embedded (see Fig.~$\ref{fig:StrongCoup}(c))$.

In Equation~\eqref{eq:alphaqs} we substitute the constant background permittivity $\epsilon_B$ by a Lorentz oscillator model for the material:
\begin{equation}
\epsilon_{B}(\omega,\omega_{\mathrm{mol}}) = \epsilon_\infty - \frac{\omega_{\mathrm{mol}}^2 f_{\mathrm{mol}}}{\omega_{\mathrm{mol}}^2-\omega^2-i\Gamma_m\omega},
\label{eq:EffectiveBackgroundPermittivity}
\end{equation}
where $f_{\mathrm{mol}}$ is the dimensionless oscillator strength of the molecular material and $\Gamma_m$ the optical loss of the effective medium. In Fig.~\ref{fig:StrongCoup}(c) we calculate the spectrum for the arrray embedded in a effective medium with $f_{\mathrm{mol}} = 10^{-3}$, $\Gamma_m = 0.1$~meV and  $\omega_{\mathrm{mol}}$ ranging from $0.9\omega_{\mathrm{sp}}$ to $1.1\omega_{\mathrm{sp}}$.

Contrasting with the case with the molecules over the edge in Fig.~\ref{fig:StrongCoup}(a), here we see large polaritonic splittings for bulk and corner states too. This is due to the molecules being extended all over space and not only placed over the edges. 

Finally, we study the energy transfer in the effective medium description by adding two extra dipoles (donor and acceptor) and using the strongly-coupled dyadic Green function. In Fig.~\ref{fig:StrongCoup}(d) we plot the spectrum overlaid on top of the energy transfer enhancement. From the spectrum, we see that the energy level splitting of the molecule-edge state polaritons is reduced due to level pushing from the bulk and corner mixed states, due to the spatial overlap of the molecular system and all of the edge, bulk and corner states. However, the energy transfer enhancement via the edge states at the frequency of the polaritons is still clearly preserved. 

% Conclusions 
\section{Conclusions}

Molecular energy transfer in free space is short-ranged, while coupling to photonic modes can substantially extend the distance over which molecules may interact, with promising applications. In this paper we have illustrated how both weak and strong coupling between molecules and the topological edge states of a 2D SSH metallic nanoparticle array can provide energy transfer enhancement using a coupled-dipole model formalism. However, we note that this concept is not limited to the SSH2D model, nor to plasmonic systems. This formalism can be used to study coupling in any finite heterogeneous array of couple dipoles, such as dielectric nanoparticle systems, quantum dots, etc. 

We have given a detailed study of the dispersion and optical properties of the photonic SSH2D model with all-to-all interactions, and demonstrated that the weak-coupling of molecules to the topological edge state of this system already provides enhancement of molecular energy transfer between a single donor and single acceptor molecule, compared to that in free space. We have shown that by considering a finite ensemble of molecules at the maxima of the local density of states at the frequency of the edge states - which correspond to the positions over the particles on the edge - leads to collective coupling between the molecules and edge states. When the distance between the molecules and the array is reduced, the strong coupling regime may be accessed, leading to polaritonic Rabi splitting of the edge states -  even for those edge states that are off-resonance. This is due to the overlap between two edge states being small but non-zero, so molecules very close to the surface may couple to states which are off-resonant. In the strong coupling regime, molecular energy transfer is enhanced by many orders of magnitude in comparison to the weak-coupled regime. 

We have verified this result at realistic molecular density by considering the nanoparticle array to be embedded in an effective medium representing the molecular material. In this case, we have shown that it is harder to couple exclusively to edge states, due to the spatial overlap of the molecular material with all photonic states. However, we have shown that coupling to edge states in this regime still leads to dramatic energy transfer enhancement. There is a great research effort currently underway to understand and control the strong coupling of molecules to photonic modes, particularly in open systems, and in order to engineer long-range molecular energy transfer. We hope that the results we have reported here will contribute to both of these efforts via the theoretical demonstration of topological light-molecule strong coupling, with various potential applications and new research avenues extending beyond the system studied in the current work. 

\begin{acknowledgement}
We thank William Wardley and Felipe Herrera for fruitful discussions. V.G. thanks the
ENSEMBLE3-Centre of Excellence for nanophotonics,
advanced materials and novel crystal growth-based technologies, project (GA No. MAB/2020/14) carried out within the International Research Agendas program of the Foundation for Polish Science cofinanced by the European Union under the
European Regional Development Fund.

A.B. and J.A.S.G. acknowledge financial support from the grants BICPLAN6G (TED2021-131417B-I00) and LIGHTCOMPAS (PID2022-137569NB-C41), funded by MCIU/AEI/10.13039/501100011033, “ERDF A way of making Europe”, and European Union NextGenerationEU/PRTR and from MCIU through predoctoral fellowship PRE2019-090689. 
\end{acknowledgement}

%\bibliographystyle{...}
%\bibliography{...}

\bibliographystyle{IEEEtran}
\bibliography{bibcoup}

\providecommand{\latin}[1]{#1}
\makeatletter
\providecommand{\doi}
  {\begingroup\let\do\@makeother\dospecials
  \catcode`\{=1 \catcode`\}=2 \doi@aux}
\providecommand{\doi@aux}[1]{\endgroup\texttt{#1}}
\makeatother
\providecommand*\mcitethebibliography{\thebibliography}
\csname @ifundefined\endcsname{endmcitethebibliography}  {\let\endmcitethebibliography\endthebibliography}{}
\begin{mcitethebibliography}{56}
\providecommand*\natexlab[1]{#1}
\providecommand*\mciteSetBstSublistMode[1]{}
\providecommand*\mciteSetBstMaxWidthForm[2]{}
\providecommand*\mciteBstWouldAddEndPuncttrue
  {\def\EndOfBibitem{\unskip.}}
\providecommand*\mciteBstWouldAddEndPunctfalse
  {\let\EndOfBibitem\relax}
\providecommand*\mciteSetBstMidEndSepPunct[3]{}
\providecommand*\mciteSetBstSublistLabelBeginEnd[3]{}
\providecommand*\EndOfBibitem{}
\mciteSetBstSublistMode{f}
\mciteSetBstMaxWidthForm{subitem}{(\alph{mcitesubitemcount})}
\mciteSetBstSublistLabelBeginEnd
  {\mcitemaxwidthsubitemform\space}
  {\relax}
  {\relax}

\bibitem[Törmä and Barnes(2015)Törmä, and Barnes]{Torma2015}
Törmä,~P.; Barnes,~W.~L. Strong coupling between surface plasmon polaritons and emitters: a review. \emph{Reports on Progress in Physics} \textbf{2015}, \emph{78}, 013901\relax
\mciteBstWouldAddEndPuncttrue
\mciteSetBstMidEndSepPunct{\mcitedefaultmidpunct}
{\mcitedefaultendpunct}{\mcitedefaultseppunct}\relax
\EndOfBibitem
\bibitem[Ribeiro \latin{et~al.}(2018)Ribeiro, Mart{\'\i}nez-Mart{\'\i}nez, Du, Campos-Gonzalez-Angulo, and Yuen-Zhou]{ribeiro2018polariton}
Ribeiro,~R.~F.; Mart{\'\i}nez-Mart{\'\i}nez,~L.~A.; Du,~M.; Campos-Gonzalez-Angulo,~J.; Yuen-Zhou,~J. Polariton chemistry: controlling molecular dynamics with optical cavities. \emph{Chemical science} \textbf{2018}, \emph{9}, 6325--6339\relax
\mciteBstWouldAddEndPuncttrue
\mciteSetBstMidEndSepPunct{\mcitedefaultmidpunct}
{\mcitedefaultendpunct}{\mcitedefaultseppunct}\relax
\EndOfBibitem
\bibitem[Dovzhenko \latin{et~al.}(2018)Dovzhenko, Ryabchuk, Rakovich, and Nabiev]{dovzhenko2018light}
Dovzhenko,~D.; Ryabchuk,~S.; Rakovich,~Y.~P.; Nabiev,~I. Light--matter interaction in the strong coupling regime: configurations, conditions, and applications. \emph{Nanoscale} \textbf{2018}, \emph{10}, 3589--3605\relax
\mciteBstWouldAddEndPuncttrue
\mciteSetBstMidEndSepPunct{\mcitedefaultmidpunct}
{\mcitedefaultendpunct}{\mcitedefaultseppunct}\relax
\EndOfBibitem
\bibitem[Nikolis \latin{et~al.}(2019)Nikolis, Mischok, Siegmund, Kublitski, Jia, Benduhn, H{\"o}rmann, Neher, Gather, Spoltore, \latin{et~al.} others]{nikolis2019strong}
Nikolis,~V.~C.; Mischok,~A.; Siegmund,~B.; Kublitski,~J.; Jia,~X.; Benduhn,~J.; H{\"o}rmann,~U.; Neher,~D.; Gather,~M.~C.; Spoltore,~D.; others Strong light-matter coupling for reduced photon energy losses in organic photovoltaics. \emph{Nature communications} \textbf{2019}, \emph{10}, 3706\relax
\mciteBstWouldAddEndPuncttrue
\mciteSetBstMidEndSepPunct{\mcitedefaultmidpunct}
{\mcitedefaultendpunct}{\mcitedefaultseppunct}\relax
\EndOfBibitem
\bibitem[Herrera and Owrutsky(2020)Herrera, and Owrutsky]{herrera2020molecular}
Herrera,~F.; Owrutsky,~J. Molecular polaritons for controlling chemistry with quantum optics. \emph{The Journal of chemical physics} \textbf{2020}, \emph{152}, 100902\relax
\mciteBstWouldAddEndPuncttrue
\mciteSetBstMidEndSepPunct{\mcitedefaultmidpunct}
{\mcitedefaultendpunct}{\mcitedefaultseppunct}\relax
\EndOfBibitem
\bibitem[Polak \latin{et~al.}(2020)Polak, Jayaprakash, Lyons, Mart{\'\i}nez-Mart{\'\i}nez, Leventis, Fallon, Coulthard, Bossanyi, Georgiou, Petty, \latin{et~al.} others]{polak2020manipulating}
Polak,~D.; Jayaprakash,~R.; Lyons,~T.~P.; Mart{\'\i}nez-Mart{\'\i}nez,~L.~{\'A}.; Leventis,~A.; Fallon,~K.~J.; Coulthard,~H.; Bossanyi,~D.~G.; Georgiou,~K.; Petty,~A.~J.; others Manipulating molecules with strong coupling: harvesting triplet excitons in organic exciton microcavities. \emph{Chemical science} \textbf{2020}, \emph{11}, 343--354\relax
\mciteBstWouldAddEndPuncttrue
\mciteSetBstMidEndSepPunct{\mcitedefaultmidpunct}
{\mcitedefaultendpunct}{\mcitedefaultseppunct}\relax
\EndOfBibitem
\bibitem[Hirai \latin{et~al.}(2023)Hirai, Hutchison, and Uji-i]{hirai2023molecular}
Hirai,~K.; Hutchison,~J.~A.; Uji-i,~H. Molecular chemistry in cavity strong coupling. \emph{Chemical Reviews} \textbf{2023}, \emph{123}, 8099--8126\relax
\mciteBstWouldAddEndPuncttrue
\mciteSetBstMidEndSepPunct{\mcitedefaultmidpunct}
{\mcitedefaultendpunct}{\mcitedefaultseppunct}\relax
\EndOfBibitem
\bibitem[Ebbesen \latin{et~al.}(2023)Ebbesen, Rubio, and Scholes]{ebbesen2023introduction}
Ebbesen,~T.~W.; Rubio,~A.; Scholes,~G.~D. Introduction: Polaritonic Chemistry. 2023\relax
\mciteBstWouldAddEndPuncttrue
\mciteSetBstMidEndSepPunct{\mcitedefaultmidpunct}
{\mcitedefaultendpunct}{\mcitedefaultseppunct}\relax
\EndOfBibitem
\bibitem[Feist and Garcia-Vidal(2015)Feist, and Garcia-Vidal]{feist2015extraordinary}
Feist,~J.; Garcia-Vidal,~F.~J. Extraordinary exciton conductance induced by strong coupling. \emph{Physical review letters} \textbf{2015}, \emph{114}, 196402\relax
\mciteBstWouldAddEndPuncttrue
\mciteSetBstMidEndSepPunct{\mcitedefaultmidpunct}
{\mcitedefaultendpunct}{\mcitedefaultseppunct}\relax
\EndOfBibitem
\bibitem[Du \latin{et~al.}(2018)Du, Martínez-Martínez, Ribeiro, Hu, Menon, and Yuen-Zhou]{Du2018}
Du,~M.; Martínez-Martínez,~L.~A.; Ribeiro,~R.~F.; Hu,~Z.; Menon,~V.~M.; Yuen-Zhou,~J. Theory for polariton-assisted remote energy transfer. \emph{Chemical Science} \textbf{2018}, \emph{9}, 6659--6669\relax
\mciteBstWouldAddEndPuncttrue
\mciteSetBstMidEndSepPunct{\mcitedefaultmidpunct}
{\mcitedefaultendpunct}{\mcitedefaultseppunct}\relax
\EndOfBibitem
\bibitem[Georgiou \latin{et~al.}(2021)Georgiou, Jayaprakash, Othonos, and Lidzey]{georgiou2021ultralong}
Georgiou,~K.; Jayaprakash,~R.; Othonos,~A.; Lidzey,~D.~G. Ultralong-Range Polariton-Assisted Energy Transfer in Organic Microcavities. \emph{Angewandte Chemie} \textbf{2021}, \emph{133}, 16797--16803\relax
\mciteBstWouldAddEndPuncttrue
\mciteSetBstMidEndSepPunct{\mcitedefaultmidpunct}
{\mcitedefaultendpunct}{\mcitedefaultseppunct}\relax
\EndOfBibitem
\bibitem[Balasubrahmaniyam \latin{et~al.}(2022)Balasubrahmaniyam, Simkovich, Golombek, Ankonina, and Schwartz]{balasubrahmaniyam2022unveiling}
Balasubrahmaniyam,~M.; Simkovich,~A.; Golombek,~A.; Ankonina,~G.; Schwartz,~T. Unveiling the mixed nature of polaritonic transport: From enhanced diffusion to ballistic motion approaching the speed of light. \emph{arXiv preprint arXiv:2205.06683} \textbf{2022}, \relax
\mciteBstWouldAddEndPunctfalse
\mciteSetBstMidEndSepPunct{\mcitedefaultmidpunct}
{}{\mcitedefaultseppunct}\relax
\EndOfBibitem
\bibitem[Jin \latin{et~al.}(2023)Jin, Sample, Sun, Gao, Deng, Li, Dou, Odom, and Huang]{jin2023enhanced}
Jin,~L.; Sample,~A.~D.; Sun,~D.; Gao,~Y.; Deng,~S.; Li,~R.; Dou,~L.; Odom,~T.~W.; Huang,~L. Enhanced Two-Dimensional Exciton Propagation via Strong Light--Matter Coupling with Surface Lattice Plasmons. \emph{ACS Photonics} \textbf{2023}, 1983–1991\relax
\mciteBstWouldAddEndPuncttrue
\mciteSetBstMidEndSepPunct{\mcitedefaultmidpunct}
{\mcitedefaultendpunct}{\mcitedefaultseppunct}\relax
\EndOfBibitem
\bibitem[Engelhardt and Cao(2023)Engelhardt, and Cao]{engelhardt2023polariton}
Engelhardt,~G.; Cao,~J. Polariton Localization and Dispersion Properties of Disordered Quantum Emitters in Multimode Microcavities. \emph{Physical Review Letters} \textbf{2023}, \emph{130}, 213602\relax
\mciteBstWouldAddEndPuncttrue
\mciteSetBstMidEndSepPunct{\mcitedefaultmidpunct}
{\mcitedefaultendpunct}{\mcitedefaultseppunct}\relax
\EndOfBibitem
\bibitem[Rider \latin{et~al.}(2019)Rider, Palmer, Pocock, Xiao, Huidobro, and Giannini]{Rider2019}
Rider,~M.~S.; Palmer,~S.~J.; Pocock,~S.~R.; Xiao,~X.; Huidobro,~P.~A.; Giannini,~V. A perspective on topological nanophotonics: Current status and future challenges. \emph{Journal of Applied Physics} \textbf{2019}, \emph{125}\relax
\mciteBstWouldAddEndPuncttrue
\mciteSetBstMidEndSepPunct{\mcitedefaultmidpunct}
{\mcitedefaultendpunct}{\mcitedefaultseppunct}\relax
\EndOfBibitem
\bibitem[Blanco~de Paz \latin{et~al.}(2019)Blanco~de Paz, Devescovi, Giedke, Saenz, Vergniory, Bradlyn, Bercioux, and Garcia-Etxarri]{BlancodePaz2019}
Blanco~de Paz,~M.; Devescovi,~C.; Giedke,~G.; Saenz,~J.~J.; Vergniory,~M.~G.; Bradlyn,~B.; Bercioux,~D.; Garcia-Etxarri,~A. {Tutorial: Computing Topological Invariants in 2D Photonic Crystals}. \emph{Advanced Quantum Technologies} \textbf{2019}, 1900117\relax
\mciteBstWouldAddEndPuncttrue
\mciteSetBstMidEndSepPunct{\mcitedefaultmidpunct}
{\mcitedefaultendpunct}{\mcitedefaultseppunct}\relax
\EndOfBibitem
\bibitem[Lee \latin{et~al.}(2023)Lee, Kim, and Park]{Lee2023}
Lee,~Y.-M.; Kim,~S.-E.; Park,~J.-E. Strong coupling in plasmonic metal nanoparticles. \emph{Nano Convergence} \textbf{2023}, \emph{10}\relax
\mciteBstWouldAddEndPuncttrue
\mciteSetBstMidEndSepPunct{\mcitedefaultmidpunct}
{\mcitedefaultendpunct}{\mcitedefaultseppunct}\relax
\EndOfBibitem
\bibitem[Hamza \latin{et~al.}(2023)Hamza, Al-Dulaimi, Bouillard, and Adawi]{Hamza2023}
Hamza,~A.~O.; Al-Dulaimi,~A.; Bouillard,~J.-S.~G.; Adawi,~A.~M. Long-Range and High-Efficiency Plasmon-Assisted F\"{o}rster Resonance Energy Transfer. \emph{The Journal of Physical Chemistry C} \textbf{2023}, \emph{127}, 21611–21616\relax
\mciteBstWouldAddEndPuncttrue
\mciteSetBstMidEndSepPunct{\mcitedefaultmidpunct}
{\mcitedefaultendpunct}{\mcitedefaultseppunct}\relax
\EndOfBibitem
\bibitem[Bitton and Haran(2022)Bitton, and Haran]{Bitton2022}
Bitton,~O.; Haran,~G. Plasmonic Cavities and Individual Quantum Emitters in the Strong Coupling Limit. \emph{Accounts of Chemical Research} \textbf{2022}, \emph{55}, 1659--1668\relax
\mciteBstWouldAddEndPuncttrue
\mciteSetBstMidEndSepPunct{\mcitedefaultmidpunct}
{\mcitedefaultendpunct}{\mcitedefaultseppunct}\relax
\EndOfBibitem
\bibitem[Sánchez-Barquilla \latin{et~al.}(2022)Sánchez-Barquilla, Fernández-Domínguez, Feist, and García-Vidal]{SanchezBarquilla2022}
Sánchez-Barquilla,~M.; Fernández-Domínguez,~A.~I.; Feist,~J.; García-Vidal,~F.~J. A Theoretical Perspective on Molecular Polaritonics. \emph{ACS Photonics} \textbf{2022}, \emph{9}, 1830–1841\relax
\mciteBstWouldAddEndPuncttrue
\mciteSetBstMidEndSepPunct{\mcitedefaultmidpunct}
{\mcitedefaultendpunct}{\mcitedefaultseppunct}\relax
\EndOfBibitem
\bibitem[Rider \latin{et~al.}(2022)Rider, Buend{\'i}a, Abujetas, Huidobro, S{\'a}nchez-Gil, and Giannini]{Rider2022}
Rider,~M.~S.; Buend{\'i}a,~{\'A}.; Abujetas,~D.~R.; Huidobro,~P.~A.; S{\'a}nchez-Gil,~J.~A.; Giannini,~V. Advances and Prospects in Topological Nanoparticle Photonics. \emph{ACS Photonics} \textbf{2022}, \emph{9}, 1483--1499\relax
\mciteBstWouldAddEndPuncttrue
\mciteSetBstMidEndSepPunct{\mcitedefaultmidpunct}
{\mcitedefaultendpunct}{\mcitedefaultseppunct}\relax
\EndOfBibitem
\bibitem[Väkeväinen \latin{et~al.}(2014)Väkeväinen, Moerland, Rekola, Eskelinen, Martikainen, Kim, and Törmä]{Vakevainen2014}
Väkeväinen,~A.~I.; Moerland,~R.~J.; Rekola,~H.~T.; Eskelinen,~A.-P.; Martikainen,~J.-P.; Kim,~D.-H.; Törmä,~P. Plasmonic Surface Lattice Resonances at the Strong Coupling Regime. \emph{Nano Letters} \textbf{2014}, \emph{14}, 1721--1727\relax
\mciteBstWouldAddEndPuncttrue
\mciteSetBstMidEndSepPunct{\mcitedefaultmidpunct}
{\mcitedefaultendpunct}{\mcitedefaultseppunct}\relax
\EndOfBibitem
\bibitem[Allard and Weick(2021)Allard, and Weick]{Allard2021}
Allard,~T.~F.; Weick,~G. Quantum theory of plasmon polaritons in chains of metallic nanoparticles: From near: From far-field coupling regime. \emph{Physical Review B} \textbf{2021}, \emph{104}, 125434\relax
\mciteBstWouldAddEndPuncttrue
\mciteSetBstMidEndSepPunct{\mcitedefaultmidpunct}
{\mcitedefaultendpunct}{\mcitedefaultseppunct}\relax
\EndOfBibitem
\bibitem[Allard and Weick(2022)Allard, and Weick]{Allard2022}
Allard,~T.~F.; Weick,~G. Disorder-enhanced transport in a chain of lossy dipoles strongly coupled to cavity photons. \emph{Phys. Rev. B} \textbf{2022}, \emph{106}, 245424\relax
\mciteBstWouldAddEndPuncttrue
\mciteSetBstMidEndSepPunct{\mcitedefaultmidpunct}
{\mcitedefaultendpunct}{\mcitedefaultseppunct}\relax
\EndOfBibitem
\bibitem[Allard and Weick(2023)Allard, and Weick]{Allard2023}
Allard,~T.~F.; Weick,~G. Multiple polaritonic edge states in a Su-Schrieffer-Heeger chain strongly coupled to a multimode cavity. \emph{Phys. Rev. B} \textbf{2023}, \emph{108}, 245417\relax
\mciteBstWouldAddEndPuncttrue
\mciteSetBstMidEndSepPunct{\mcitedefaultmidpunct}
{\mcitedefaultendpunct}{\mcitedefaultseppunct}\relax
\EndOfBibitem
\bibitem[Thanopulos \latin{et~al.}(2023)Thanopulos, Yannopapas, and Paspalakis]{Thanopulos2023}
Thanopulos,~I.; Yannopapas,~V.; Paspalakis,~E. Strong Coupling Dynamics of a Quantum Emitter near a Topological Insulator Nanoparticle. \emph{Nanomaterials} \textbf{2023}, \emph{13}, 2787\relax
\mciteBstWouldAddEndPuncttrue
\mciteSetBstMidEndSepPunct{\mcitedefaultmidpunct}
{\mcitedefaultendpunct}{\mcitedefaultseppunct}\relax
\EndOfBibitem
\bibitem[Barik \latin{et~al.}(2018)Barik, Karasahin, Flower, Cai, Miyake, DeGottardi, Hafezi, and Waks]{barik2018topological}
Barik,~S.; Karasahin,~A.; Flower,~C.; Cai,~T.; Miyake,~H.; DeGottardi,~W.; Hafezi,~M.; Waks,~E. A topological quantum optics interface. \emph{Science} \textbf{2018}, \emph{359}, 666--668\relax
\mciteBstWouldAddEndPuncttrue
\mciteSetBstMidEndSepPunct{\mcitedefaultmidpunct}
{\mcitedefaultendpunct}{\mcitedefaultseppunct}\relax
\EndOfBibitem
\bibitem[Kim and Rho(2023)Kim, and Rho]{Kim2023}
Kim,~M.; Rho,~J. CDPDS: Coupled dipole method-based photonic dispersion solver. \emph{Computer Physics Communications} \textbf{2023}, \emph{282}, 108493\relax
\mciteBstWouldAddEndPuncttrue
\mciteSetBstMidEndSepPunct{\mcitedefaultmidpunct}
{\mcitedefaultendpunct}{\mcitedefaultseppunct}\relax
\EndOfBibitem
\bibitem[Su \latin{et~al.}(1979)Su, Schrieffer, and Heeger]{su1979solitons}
Su,~W.-P.; Schrieffer,~J.~R.; Heeger,~A.~J. Solitons in polyacetylene. \emph{Physical review letters} \textbf{1979}, \emph{42}, 1698\relax
\mciteBstWouldAddEndPuncttrue
\mciteSetBstMidEndSepPunct{\mcitedefaultmidpunct}
{\mcitedefaultendpunct}{\mcitedefaultseppunct}\relax
\EndOfBibitem
\bibitem[Baraclough \latin{et~al.}(2019)Baraclough, Seetharaman, Hooper, and Barnes]{Baraclough_ACSPhot_2019_6_3003}
Baraclough,~M.; Seetharaman,~S.~S.; Hooper,~I.~R.; Barnes,~W.~L. Metamaterial Analogues of Molecular Aggregates. \emph{ACS Photonics} \textbf{2019}, \emph{6}, 3003--3009\relax
\mciteBstWouldAddEndPuncttrue
\mciteSetBstMidEndSepPunct{\mcitedefaultmidpunct}
{\mcitedefaultendpunct}{\mcitedefaultseppunct}\relax
\EndOfBibitem
\bibitem[Downing and Weick(2017)Downing, and Weick]{Downing2017}
Downing,~C.~A.; Weick,~G. Topological collective plasmons in bipartite chains of metallic nanoparticles. \emph{Phys. Rev. B} \textbf{2017}, \emph{95}, 125426\relax
\mciteBstWouldAddEndPuncttrue
\mciteSetBstMidEndSepPunct{\mcitedefaultmidpunct}
{\mcitedefaultendpunct}{\mcitedefaultseppunct}\relax
\EndOfBibitem
\bibitem[Pocock \latin{et~al.}(2018)Pocock, Xiao, Huidobro, and Giannini]{Pocock2018}
Pocock,~S.~R.; Xiao,~X.; Huidobro,~P.~A.; Giannini,~V. Topological Plasmonic Chain with Retardation and Radiative Effects. \emph{ACS Photonics} \textbf{2018}, \emph{5}, 2271–2279\relax
\mciteBstWouldAddEndPuncttrue
\mciteSetBstMidEndSepPunct{\mcitedefaultmidpunct}
{\mcitedefaultendpunct}{\mcitedefaultseppunct}\relax
\EndOfBibitem
\bibitem[Buendía \latin{et~al.}(2023)Buendía, Sánchez-Gil, and Giannini]{Buendia2023}
Buendía,~A.; Sánchez-Gil,~J.~A.; Giannini,~V. Exploiting Oriented Field Projectors to Open Topological Gaps in Plasmonic Nanoparticle Arrays. \emph{ACS Photonics} \textbf{2023}, \emph{10}, 464–474\relax
\mciteBstWouldAddEndPuncttrue
\mciteSetBstMidEndSepPunct{\mcitedefaultmidpunct}
{\mcitedefaultendpunct}{\mcitedefaultseppunct}\relax
\EndOfBibitem
\bibitem[Honari-Latifpour and Yousefi(2019)Honari-Latifpour, and Yousefi]{Honari2019}
Honari-Latifpour,~M.; Yousefi,~L. Topological plasmonic edge states in a planar array of metallic nanoparticles. \emph{Nanophotonics} \textbf{2019}, \emph{8}, 799--806\relax
\mciteBstWouldAddEndPuncttrue
\mciteSetBstMidEndSepPunct{\mcitedefaultmidpunct}
{\mcitedefaultendpunct}{\mcitedefaultseppunct}\relax
\EndOfBibitem
\bibitem[Proctor \latin{et~al.}(2019)Proctor, Craster, Maier, Giannini, and Huidobro]{Proctor2019}
Proctor,~M.; Craster,~R.~V.; Maier,~S.~A.; Giannini,~V.; Huidobro,~P.~A. Exciting Pseudospin-Dependent Edge States in Plasmonic Metasurfaces. \emph{ACS Photonics} \textbf{2019}, \emph{6}, 2985–2995\relax
\mciteBstWouldAddEndPuncttrue
\mciteSetBstMidEndSepPunct{\mcitedefaultmidpunct}
{\mcitedefaultendpunct}{\mcitedefaultseppunct}\relax
\EndOfBibitem
\bibitem[Kim and Rho(2020)Kim, and Rho]{Kim2020}
Kim,~M.; Rho,~J. Topological edge and corner states in a two-dimensional photonic Su-Schrieffer-Heeger lattice. \emph{Nanophotonics} \textbf{2020}, \emph{9}, 3227–3234\relax
\mciteBstWouldAddEndPuncttrue
\mciteSetBstMidEndSepPunct{\mcitedefaultmidpunct}
{\mcitedefaultendpunct}{\mcitedefaultseppunct}\relax
\EndOfBibitem
\bibitem[Kim \latin{et~al.}(2020)Kim, Hwang, Smirnova, Jeong, Kivshar, and Park]{Kim2020lasing}
Kim,~H.-R.; Hwang,~M.-S.; Smirnova,~D.; Jeong,~K.-Y.; Kivshar,~Y.; Park,~H.-G. Multipolar lasing modes from topological corner states. \emph{Nature Communications} \textbf{2020}, \emph{11}, 5758\relax
\mciteBstWouldAddEndPuncttrue
\mciteSetBstMidEndSepPunct{\mcitedefaultmidpunct}
{\mcitedefaultendpunct}{\mcitedefaultseppunct}\relax
\EndOfBibitem
\bibitem[Wu \latin{et~al.}(2023)Wu, Ghosh, Gan, Shi, Mandal, Sun, Zhang, Liew, Su, and Xiong]{Wu2023}
Wu,~J.; Ghosh,~S.; Gan,~Y.; Shi,~Y.; Mandal,~S.; Sun,~H.; Zhang,~B.; Liew,~T. C.~H.; Su,~R.; Xiong,~Q. Higher-order topological polariton corner state lasing. \emph{Science Advances} \textbf{2023}, \emph{9}\relax
\mciteBstWouldAddEndPuncttrue
\mciteSetBstMidEndSepPunct{\mcitedefaultmidpunct}
{\mcitedefaultendpunct}{\mcitedefaultseppunct}\relax
\EndOfBibitem
\bibitem[Xie \latin{et~al.}(2018)Xie, Wang, Wang, Zhu, Jiang, Lu, and Chen]{Xie2018}
Xie,~B.-Y.; Wang,~H.-F.; Wang,~H.-X.; Zhu,~X.-Y.; Jiang,~J.-H.; Lu,~M.-H.; Chen,~Y.-F. Second-order photonic topological insulator with corner states. \emph{Phys. Rev. B} \textbf{2018}, \emph{98}, 205147\relax
\mciteBstWouldAddEndPuncttrue
\mciteSetBstMidEndSepPunct{\mcitedefaultmidpunct}
{\mcitedefaultendpunct}{\mcitedefaultseppunct}\relax
\EndOfBibitem
\bibitem[Obana \latin{et~al.}(2019)Obana, Liu, and Wakabayashi]{Obana2019}
Obana,~D.; Liu,~F.; Wakabayashi,~K. Topological edge states in the Su-Schrieffer-Heeger model. \emph{Physical Review B} \textbf{2019}, \emph{100}, 075437\relax
\mciteBstWouldAddEndPuncttrue
\mciteSetBstMidEndSepPunct{\mcitedefaultmidpunct}
{\mcitedefaultendpunct}{\mcitedefaultseppunct}\relax
\EndOfBibitem
\bibitem[Benalcazar and Cerjan(2020)Benalcazar, and Cerjan]{Benalcazar2020}
Benalcazar,~W.~A.; Cerjan,~A. Bound states in the continuum of higher-order topological insulators. \emph{Phys. Rev. B} \textbf{2020}, \emph{101}, 161116\relax
\mciteBstWouldAddEndPuncttrue
\mciteSetBstMidEndSepPunct{\mcitedefaultmidpunct}
{\mcitedefaultendpunct}{\mcitedefaultseppunct}\relax
\EndOfBibitem
\bibitem[Schlömer \latin{et~al.}(2021)Schlömer, Jiang, and Haas]{Schlomer2021}
Schlömer,~H.; Jiang,~Z.; Haas,~S. Plasmons in two-dimensional topological insulators. \emph{Physical Review B} \textbf{2021}, \emph{103}, 115116\relax
\mciteBstWouldAddEndPuncttrue
\mciteSetBstMidEndSepPunct{\mcitedefaultmidpunct}
{\mcitedefaultendpunct}{\mcitedefaultseppunct}\relax
\EndOfBibitem
\bibitem[Luo \latin{et~al.}(2023)Luo, Li, Shen, and Deng]{Luo2023}
Luo,~C.; Li,~H.-C.; Shen,~Y.; Deng,~X.-H. Topological robust corner states of a two-dimensional square lattice with C4 symmetry in fully coupled dipolar arrays. \emph{Europhysics Letters} \textbf{2023}, \emph{143}, 56002\relax
\mciteBstWouldAddEndPuncttrue
\mciteSetBstMidEndSepPunct{\mcitedefaultmidpunct}
{\mcitedefaultendpunct}{\mcitedefaultseppunct}\relax
\EndOfBibitem
\bibitem[Liu \latin{et~al.}(2019)Liu, Gao, Zhang, Ma, Zhang, Liu, Xiang, Cui, and Zhang]{Liu2019}
Liu,~S.; Gao,~W.; Zhang,~Q.; Ma,~S.; Zhang,~L.; Liu,~C.; Xiang,~Y.~J.; Cui,~T.~J.; Zhang,~S. Topologically Protected Edge State in Two-Dimensional Su–Schrieffer–Heeger Circuit. \emph{Research} \textbf{2019}, \emph{2019}\relax
\mciteBstWouldAddEndPuncttrue
\mciteSetBstMidEndSepPunct{\mcitedefaultmidpunct}
{\mcitedefaultendpunct}{\mcitedefaultseppunct}\relax
\EndOfBibitem
\bibitem[Hu \latin{et~al.}(2021)Hu, Bongiovanni, Juki{\'{c}}, Jajti{\'{c}}, Xia, Song, Xu, Morandotti, Buljan, and Chen]{Hu2021}
Hu,~Z.; Bongiovanni,~D.; Juki{\'{c}},~D.; Jajti{\'{c}},~E.; Xia,~S.; Song,~D.; Xu,~J.; Morandotti,~R.; Buljan,~H.; Chen,~Z. Nonlinear control of photonic higher-order topological bound states in the continuum. \emph{Light: Science \& Applications} \textbf{2021}, \emph{10}, 8609875\relax
\mciteBstWouldAddEndPuncttrue
\mciteSetBstMidEndSepPunct{\mcitedefaultmidpunct}
{\mcitedefaultendpunct}{\mcitedefaultseppunct}\relax
\EndOfBibitem
\bibitem[Heilmann \latin{et~al.}(2022)Heilmann, Salerno, Cuerda, Hakala, and T\"{o}rm\"{a}]{Heilmann2022}
Heilmann,~R.; Salerno,~G.; Cuerda,~J.; Hakala,~T.~K.; T\"{o}rm\"{a},~P. Quasi-{BIC} Mode Lasing in a Quadrumer Plasmonic Lattice. \emph{{ACS} Photonics} \textbf{2022}, \emph{9}, 224--232\relax
\mciteBstWouldAddEndPuncttrue
\mciteSetBstMidEndSepPunct{\mcitedefaultmidpunct}
{\mcitedefaultendpunct}{\mcitedefaultseppunct}\relax
\EndOfBibitem
\bibitem[Goerlitzer \latin{et~al.}(2023)Goerlitzer, Zapata-Herrera, Ponomareva, Feller, Garcia-Etxarri, Karg, Aizpurua, and Vogel]{Goerlitzer2023}
Goerlitzer,~E. S.~A.; Zapata-Herrera,~M.; Ponomareva,~E.; Feller,~D.; Garcia-Etxarri,~A.; Karg,~M.; Aizpurua,~J.; Vogel,~N. Molecular-Induced Chirality Transfer to Plasmonic Lattice Modes. \emph{ACS Photonics} \textbf{2023}, \emph{10}, 1821–1831\relax
\mciteBstWouldAddEndPuncttrue
\mciteSetBstMidEndSepPunct{\mcitedefaultmidpunct}
{\mcitedefaultendpunct}{\mcitedefaultseppunct}\relax
\EndOfBibitem
\bibitem[Xiao \latin{et~al.}(2023)Xiao, Gillibert, Foti, Coulon, Ulysse, Levato, Maier, Giannini, Gucciardi, and Rizza]{Xiao2023}
Xiao,~X.; Gillibert,~R.; Foti,~A.; Coulon,~P.~E.; Ulysse,~C.; Levato,~T.; Maier,~S.~A.; Giannini,~V.; Gucciardi,~P.~G.; Rizza,~G. Plasmonic Polarization Rotation in SERS Spectroscopy. \emph{Nano Letters} \textbf{2023}, \emph{23}, 2530--2535\relax
\mciteBstWouldAddEndPuncttrue
\mciteSetBstMidEndSepPunct{\mcitedefaultmidpunct}
{\mcitedefaultendpunct}{\mcitedefaultseppunct}\relax
\EndOfBibitem
\bibitem[Cherqui \latin{et~al.}(2019)Cherqui, Bourgeois, Wang, and Schatz]{Cherqui2019}
Cherqui,~C.; Bourgeois,~M.~R.; Wang,~D.; Schatz,~G.~C. Plasmonic Surface Lattice Resonances: Theory and Computation. \emph{Accounts of Chemical Research} \textbf{2019}, \emph{52}, 2548–2558\relax
\mciteBstWouldAddEndPuncttrue
\mciteSetBstMidEndSepPunct{\mcitedefaultmidpunct}
{\mcitedefaultendpunct}{\mcitedefaultseppunct}\relax
\EndOfBibitem
\bibitem[Zundel \latin{et~al.}(2022)Zundel, Cuartero-González, Sanders, Fernández-Domínguez, and Manjavacas]{Zundel2022}
Zundel,~L.; Cuartero-González,~A.; Sanders,~S.; Fernández-Domínguez,~A.~I.; Manjavacas,~A. Green Tensor Analysis of Lattice Resonances in Periodic Arrays of Nanoparticles. \emph{ACS Photonics} \textbf{2022}, \emph{9}, 540–550\relax
\mciteBstWouldAddEndPuncttrue
\mciteSetBstMidEndSepPunct{\mcitedefaultmidpunct}
{\mcitedefaultendpunct}{\mcitedefaultseppunct}\relax
\EndOfBibitem
\bibitem[Proctor \latin{et~al.}(2020)Proctor, Xiao, Craster, Maier, Giannini, and Arroyo~Huidobro]{Proctor2020}
Proctor,~M.; Xiao,~X.; Craster,~R.~V.; Maier,~S.~A.; Giannini,~V.; Arroyo~Huidobro,~P. Near- and Far-Field Excitation of Topological Plasmonic Metasurfaces. \emph{Photonics} \textbf{2020}, \emph{7}, 81\relax
\mciteBstWouldAddEndPuncttrue
\mciteSetBstMidEndSepPunct{\mcitedefaultmidpunct}
{\mcitedefaultendpunct}{\mcitedefaultseppunct}\relax
\EndOfBibitem
\bibitem[Abujetas and Sánchez-Gil(2021)Abujetas, and Sánchez-Gil]{Abujetas2021}
Abujetas,~D.~R.; Sánchez-Gil,~J.~A. Near-field excitation of bound states in the continuum in all-dielectric metasurfaces through a coupled electric/magnetic dipole model. \emph{Nanomaterials} \textbf{2021}, \emph{11}, 998\relax
\mciteBstWouldAddEndPuncttrue
\mciteSetBstMidEndSepPunct{\mcitedefaultmidpunct}
{\mcitedefaultendpunct}{\mcitedefaultseppunct}\relax
\EndOfBibitem
\bibitem[Lunnemann and Koenderink(2016)Lunnemann, and Koenderink]{Lunnemann2016}
Lunnemann,~P.; Koenderink,~A.~F. The local density of optical states of a metasurface. \emph{Scientific Reports} \textbf{2016}, \emph{6}, 20655\relax
\mciteBstWouldAddEndPuncttrue
\mciteSetBstMidEndSepPunct{\mcitedefaultmidpunct}
{\mcitedefaultendpunct}{\mcitedefaultseppunct}\relax
\EndOfBibitem
\bibitem[Novotny and Hecht(2012)Novotny, and Hecht]{Novotny2012}
Novotny,~L.; Hecht,~B. \emph{Principles of Nano-Optics}; Cambridge University Press, 2012\relax
\mciteBstWouldAddEndPuncttrue
\mciteSetBstMidEndSepPunct{\mcitedefaultmidpunct}
{\mcitedefaultendpunct}{\mcitedefaultseppunct}\relax
\EndOfBibitem
\bibitem[Ebbesen(2016)]{Ebbesen_ACS_Accounts_2016_49_2403}
Ebbesen,~T.~W. Hybrid Light--Matter States in a Molecular and Material Science Perspective. \emph{Accounts of Chemical Research} \textbf{2016}, \emph{49}, 2403--2412\relax
\mciteBstWouldAddEndPuncttrue
\mciteSetBstMidEndSepPunct{\mcitedefaultmidpunct}
{\mcitedefaultendpunct}{\mcitedefaultseppunct}\relax
\EndOfBibitem
\end{mcitethebibliography}

\appendix
\counterwithin{figure}{section}
\counterwithin{equation}{section}

\section{Normal modes and symmetries}
\label{sec:NormalModeSymmetries}
Even when finite SSH2D arrays break periodicity, when they are commensurate with the unit cell and respect the spatial symmetries, normal modes still inherit properties from the periodic system.

The bulk modes respect the symmetries of the bands of the infinite system, this is the dipolar moments within a single unit cell are either symmetric or antisymmetric in $x$ and $y$. Depending on parities in $x$ and $y$, we label the bands as $B_1 (--), B_2 (-+), B_3 (+-), B_4 (++)$. $B_4$ band is the brightest and highest frequency one. Due to rotational symmetry, $B_2$ and $B_3$ bands are degenerate. The $B_1$ band is the lowest and darkest band. 

The edge modes respect the symmetries of bulk modes in SSH chains. Within dimers, modes are either symmetric (bright) or antisymmetric (dark). Bright ($E_1-E_2$) and dark ($E_3-E_4$) modes are doubly degenerate, depending on whether they are symmetric or antisymmetric by a 90º rotation. In Fig.~\ref{fig:SSH2D}(c) we plot the main contributions to $E_1,E_3$ modes at $\Gamma$ point. This corresponds to the modes where the dimers along a single edge are in-phase. 

Corner modes ($C_1-C_4$) are quadruply degenerated and pinned to $\omega_{sp}$ due to the generalized sublattice symmetry. 

\section{Strongly coupled effective Green dyadic}
\label{sec: Strongly_Coupled_Green_Dyadic}

In order to derive the strongly-coupled effective Green dyadic, we first consider an incident field $\hat{\textbf{E}}_{\mathrm{inc}}$ that excites only the donor molecule $n_\mathrm{D}$, $[\hat{\textbf{E}}_{\mathrm{inc}}]_{3n-2;3n} = \textbf{E}_\mathrm{D} \delta_{n n_\mathrm{D}}$, where $\textbf{E}_\mathrm{D} = \alpha_{\mathrm{mol}}^{-1}(\omega)\textbf{p}_\mathrm{D}$, being $\textbf{p}_\mathrm{D}$ the electric dipole induced in the donor. The response of the molecular-plasmonic array to this excitation is given by $\textbf{P} = \alpha_{\mathrm{eff}}^{\mathrm{NP}-\mathrm{mol}}(\omega) \hat{\textbf{E}}_{\mathrm{inc}}$. Then, the dipole induced in the acceptor molecule $n_\mathrm{A}$ is given by:

\begin{eqnarray}
\textbf{p}_A = \left[\alpha_{\mathrm{eff}}^{\mathrm{NP}-\mathrm{mol}}(\omega)\hat{\textbf{E}}_{\mathrm{inc}}\right]_{3n_\mathrm{A}-2;3n_\mathrm{A}} = \left[\alpha_{\mathrm{eff}}^{\mathrm{NP}-\mathrm{mol}}(\omega)\right]_{3n_\mathrm{A}-2;3n_\mathrm{A},3n_\mathrm{D}-2;3n_\mathrm{D}} 
\overleftrightarrow{\alpha}_{\mathrm{mol}}^{-1}(\omega) \textbf{p}_\mathrm{D}
\end{eqnarray}

By analogy with the dipole induced by another dipole in free-space (from Eq.~\ref{eq:coupleddip}) $\textbf{p}_A = \frac{k^2}{\epsilon_0}\overleftrightarrow{\alpha}_{mol}(\omega) \overleftrightarrow{\textbf{G}}(\omega,\textbf{r}_A,\textbf{r}_D)\textbf{p}_D$, we can define an effective Green dyadic via strong coupling to the nanoparticle-molecular ensemble as: 

\begin{eqnarray}
\overleftrightarrow{\textbf{G}}_{\mathrm{eff}}^{\mathrm{SC}}(\omega,\textbf{r}_\mathrm{D}, \textbf{r}_\mathrm{A})   = \frac{\epsilon_0}{k^2} \overleftrightarrow{\alpha}_{\mathrm{mol}}^{-1}(\omega)
\left[\overleftrightarrow{\alpha}_{\mathrm{eff}}^{\mathrm{NP}\mbox{-}\mathrm{mol}}(\omega)\right]_{3n_\mathrm{D}-2;3n_\mathrm{D},3n_\mathrm{A}-2;3n_\mathrm{A}}  \overleftrightarrow{\alpha}_{\mathrm{mol}}^{-1}(\omega)
\label{eq:SCEffGreen_2}
\end{eqnarray}

In the uncoupled limit $z \rightarrow \infty$ and when the distance between molecules is large enough that $\alpha^{-1}_{\mathrm{mol}}(\omega) \gg  G_{zz}(\omega, \textbf{r}_\mathrm{D}, \textbf{r}_\mathrm{A})$, we recover the free-space dipole-dipole Green function: 

\begin{eqnarray} 
\lim_{z\rightarrow \infty} [\overleftrightarrow{\textbf{G}}_{\mathrm{eff}}^{\mathrm{SC}} (\omega, \textbf{r}_\mathrm{D},\textbf{r}_\mathrm{A})]_{zz} =\frac{\epsilon}{k^2} \alpha_{\mathrm{mol}}^{-2}  (\omega) \left[\begin{pmatrix} \alpha_{\mathrm{mol}}^{-1}(\omega) & -\frac{k^2}{\epsilon_0} G_{zz}(\omega,\textbf{r}_\mathrm{D}, \textbf{r}_\mathrm{A}) \\  -\frac{k^2}{\epsilon_0} G_{zz}(\omega,\textbf{r}_\mathrm{D}, \textbf{r}_\mathrm{A}) & \alpha_{\mathrm{mol}}^{-1}(\omega)  \end{pmatrix}^{-1} \right]_{1,2} = \nonumber \\ = \frac{\epsilon_0}{k^2} \frac{\alpha_{\mathrm{mol}}^{-2}(\omega)}{\alpha_{\mathrm{mol}}^{-2}(\omega)-G_{zz}(\omega,\textbf{r}_\mathrm{D}, \textbf{r}_\mathrm{A})^2}   \left[ \begin{pmatrix} \alpha_{\mathrm{mol}}^{-1}(\omega) & \frac{k^2}{\epsilon_0} G_{zz}(\omega,\textbf{r}_\mathrm{D}, \textbf{r}_\mathrm{A}) \\  \frac{k^2}{\epsilon_0} G_{zz}(\omega,\textbf{r}_\mathrm{D}, \textbf{r}_\mathrm{A}) & \alpha_{\mathrm{mol}}^{-1}(\omega)  \end{pmatrix} \right]_{1,2} \nonumber \\ \simeq G_{zz}(\omega, \textbf{r}_\mathrm{D}, \textbf{r}_\mathrm{A}) 
\end{eqnarray}

\section{Strong coupling to dark edge states}
 \label{sec:StrongCouplingDark}
 
\begin{figure*}[h!]
\begin{center}
\includegraphics[width=0.99\textwidth]{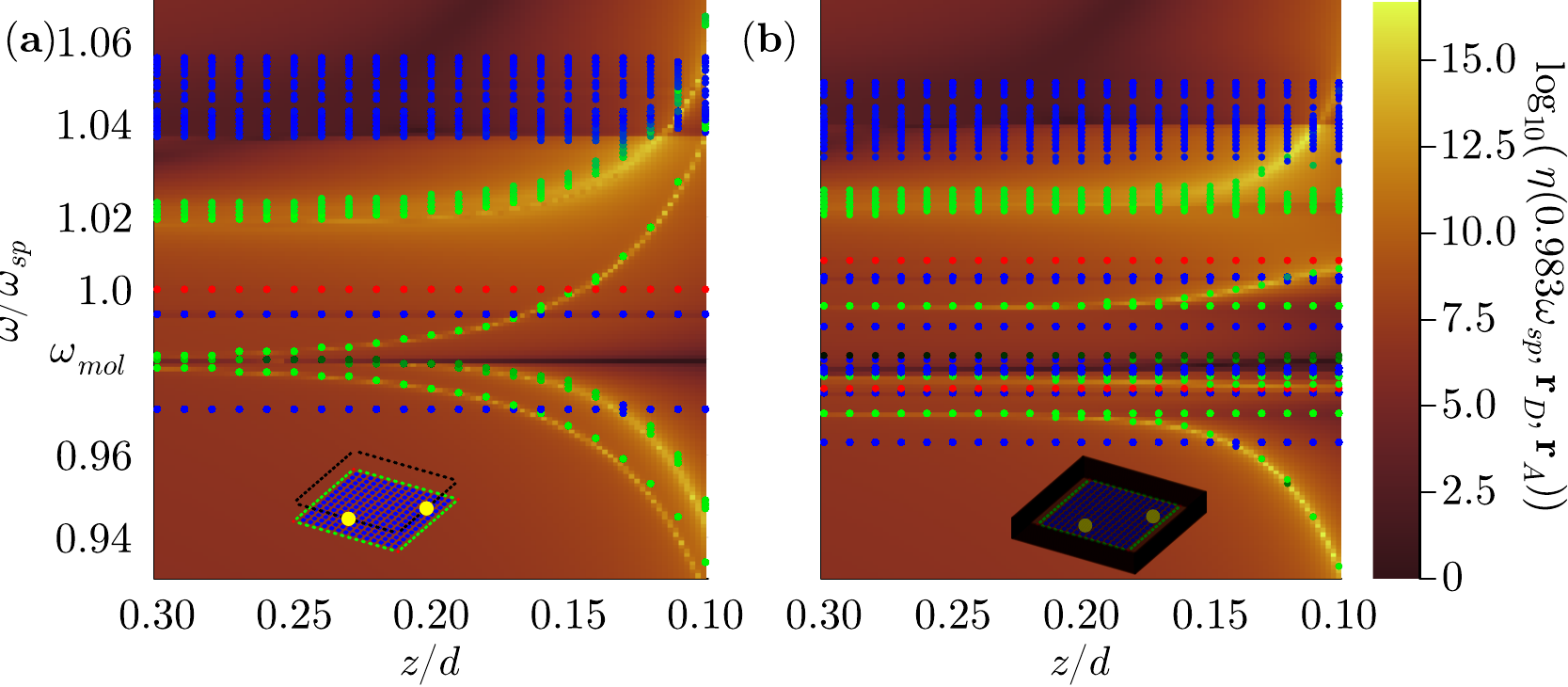}
\end{center}
\caption{\textbf{Energy transfer mediated by strong coupling between molecules and topological dark edge states}: (a) Coupling to an array of $30\times 30$ silver nanoparticles, with $a = 10$~nm, $d = 150$~nm, intra-cell distance $\beta d/2 = 0.8a$ and $\hbar\omega_{\mathrm{sp}} = 2.50$~eV and losses $\hbar\gamma = 1$~meV, by placing molecules, resonant with the dark edge states ($\omega_{\mathrm{mol}} = 0.983\omega_{\mathrm{sp}}$), over the edge nanoparticles, at the hotspots of the LDOS. Spectrum overlaid on top of enhancement of energy transfer between molecules on adjacent edges, varying the distance between the molecules and the plasmonic array $z/d$. Red, green and blue dotted lines are corner, edge and bulk states. Black dashed line is $\omega = \omega_{mol}$. Yellow dots are the donor/acceptor molecules (b) Coupling to the same array embedded in an effective molecular medium. Spectrum overlaid on top of enhancement of energy transfer between donor and acceptor molecules on adjacent edges varying with the vertical distance between the molecules and the plasmonic array $z/d$.}
\label{fig:StrongCoupDark}
\end{figure*} 

As we stated in the main text, dark states, despite their name, can be probed and excited locally by near-field coupling. However, it may be trickier to excite them, since these modes break the mirror symmetry within the unit cell, but preserve inversion invariance.

In order to couple to these modes more efficiently, we can break mirror symmetry in the excitation by placing more than one molecule per plasmonic unit cell or  with one molecule polarized in-plane. Here we restrict to $z$-polarized molecules. By placing the molecules on every hotspot (over the particles on the edge), we can couple to both dark and bright edge states. 

In Fig.~\ref{fig:StrongCoupDark}(a) we plot the spectrum and energy transfer enhancement varying the distance between nanoparticles and molecules from $z=0.3d$ to $z=0.1d$. Red, green and blue dots represent the frequencies of the corner, edge and bulk modes.

 We choose the molecules to be resonant with the dark edge states, such that $\omega_{\mathrm{mol}} = 0.983\omega_{\mathrm{sp}}$. As with the bright edge state when the molecules are very close to the metasurface ($z/d \lesssim 0.2$), they do not couple just with the edge state on-resonance but also with the off-resonance edge states, due to the small spectral overlap between the edge states.  

In Fig.~\ref{fig:StrongCoupDark}(b) we plot the spectrum and energy transfer enhancement in the effective medium description, with similars results to the coupling to bright states. We still find the larger enhancement factor are at the frequency of the (off-resonance) bright states. 

 Even when the enhancement of the energy transfer is smaller than for bright edge states, they will generally be less screened by bulk or corner states, as bright modes are broader spectrally. This may lead to more confinement in the edge in the energy transfer than for bright modes.

\end{document}